\documentclass[universe,review,accept,pdftex,oneauthor]{Definitions/mdpi} 
\firstpage{1} 
\pubvolume{9}
\issuenum{1}
\articlenumber{56}
\pubyear{2023}
\copyrightyear{2023}
\externaleditor{Academic Editors: Ivan De Martino and Mahmood Roshan}
\datereceived{30 November 2022} 
\daterevised{11 January 2023 } 
\dateaccepted{12 January 2023} 
\datepublished{16 January 2023} 
\hreflink{https://doi.org/\linebreak10.3390/universe9010056} 

\usepackage[]{trackchanges} 
\addeditor{VC}

\newcommand{\prl}{Phys. Rev. Lett.}

\newcommand{\araa}{Ann. Rev. Astron. Astrophys.}
\newcommand{\nat}{Nature}
\newcommand{\jcap}{J. Cosmol. Astropart. Phys.}
\newcommand{\prd}{Phys. Rev. D}

\Title{Dark Coincidences: Small-Scale Solutions with Refracted Gravity and MOND}

\TitleCitation{Dark Coincidences: Small-Scale Solutions with Refracted Gravity and MOND}

\Author{Valentina Cesare \orcidA{}}

\AuthorNames{Valentina Cesare}

\AuthorCitation{Cesare, V.}

\address[1]{\textls[-15]{Istituto Nazionale di Astrofisica, Astrophysical Observatory of Catania, Via Santa Sofia 78, 95123 Catania, CT, Italy;} valentina.cesare@inaf.it}

\abstract{General relativity and its Newtonian weak field limit are not sufficient to explain the observed phenomenology in the Universe, from the formation of large-scale structures to the dynamics of galaxies, with the only presence of baryonic matter. The most investigated cosmological model, the $\Lambda$CDM, accounts for the majority of observations by introducing two dark components, dark energy and dark matter, which represent $\sim$95\% of the mass-energy budget of the Universe. Nevertheless, the $\Lambda$CDM model faces important challenges on the scale of galaxies. For example, some very tight relations between the properties of dark and baryonic matters in disk galaxies, such as the baryonic Tully--Fisher relation (BTFR), the mass discrepancy--acceleration relation (MDAR), and the radial acceleration relation (RAR), which see the emergence of the acceleration scale $a_0 \simeq 1.2 \times 10^{-10}$~m~s$^{-2}$, cannot be intuitively explained by the CDM paradigm, where cosmic structures form through a stochastic merging process. An even more outstanding coincidence is due to the fact that the acceleration scale $a_0$, emerging from galaxy dynamics, also seems to be related to the cosmological constant $\Lambda$. Another challenge is provided by dwarf galaxies, which are darker than what is expected in their innermost regions. These pieces of evidence can be more naturally explained, or sometimes even predicted, by modified theories of gravity, that do not introduce any dark fluid. I illustrate possible solutions to these problems with the modified theory of gravity MOND, which departs from Newtonian gravity for accelerations smaller than $a_0$, and with Refracted Gravity, a novel classical theory of gravity introduced in 2016, where the modification of the law of gravity is instead regulated by a density scale.}


\begin{document}



\section{Introduction}
\label{sec:Intro}

If we assume the validity of standard gravity, that is, General Relativity (GR), and~the presence of ordinary (baryonic) matter alone, we cannot explain the observations from the largest to the smallest scales of the Universe. The~cosmic microwave background (CMB) radiation~\cite{Planck_2018}, the~dynamics and the large-scale distribution of the cosmic structure~\cite{Davis_1985,Springel_2006}, the~gravitational lensing, with~particular reference to the Bullet Cluster~\cite{Markevitch_2006,Clowe_2006,Paraficz_2016}, the~dynamics of galaxy clusters~\cite{Zwicky_1933}, and~the flat trend of the rotation curves of disk galaxies~\cite{Rubin_and_Ford_1970,Bosma_1978,Sanders_1990} all show a mass discrepancy of $\sim$80--90\%. Moreover, the~Hubble diagram of Ia Supernovae~\cite{Kirshner_1996,Peebles_and_Ratra_2003} proved the expansion of the Universe to be accelerated, which is not what we expect from the attractive nature of the gravity force. This probably represents the most important open question in modern cosmology and it launches several lines of~research.

With the only presence of baryonic matter, GR cannot justify $\sim$95\% of the components of the Universe. The~most investigated solution is provided by the $\Lambda$ cold dark matter ($\Lambda$CDM) cosmological model, which assumes the validity of GR and introduces two dark components besides baryonic matter to explain this missing $\sim$95\%. 

The first dark component is an exotic fluid with negative pressure called dark energy (DE), which causes the accelerated expansion of the Universe and explains $\sim$70\% of its mass-energy budget. The~DE can be identified with the cosmological constant $\Lambda$, present in the Einstein equations. The~second dark component is an invisible form of matter with a nonbaryonic nature, the~dark matter (DM), which interacts with 
baryonic matter only gravitationally and explains $\sim$25\% of the mass-energy budget of the Universe. To~account for the observed bottom-up formation scenario of cosmic structures, DM has also to be ``cold'', i.e.,~nonrelativistic at the epoch of decoupling from radiation. 

The $\Lambda$CDM model can account for almost all the mentioned pieces of evidence. Yet, it presents some issues, both on cosmological and small scales. Three of the most puzzling challenges on large scales are the cosmological constant problem~\cite{Weinberg_1989,Luo_2014}, the~coincidence problem~\cite{Velten_2014}, and~the tensions between the cosmological parameters inferred from the late and the early Universe~\cite{Planck_2018,Fleury_2013,Douspis_2019}. However, the~most important problems of the $\Lambda$CDM paradigm are observed on the scale of galaxies. The~cusp/core, missing satellites, too-big-to-fail, and~planes of satellite galaxies problems are among the most investigated  (see~\cite{DelPopolo_and_LeDelliou_2017,deMartino_2020} for a review).

Among the small-scale problems, particularly relevant is also the presence of very tight scaling relations between the properties of dark and baryonic matters in galaxies. These are the baryonic Tully--Fisher relation (BTFR)~\cite{McGaugh_BTFR_2000}, the~mass discrepancy--acceleration relation (MDAR)~\cite{McGaugh_MDAR_2004}, and~the radial acceleration relation (RAR)~\cite{McGaugh_RAR_2016}, whose scatter is consistent with the observational errors. In~a $\Lambda$CDM context, it is not trivial to explain why quantities related to DM, which represents $\sim$90\% of the galaxy mass, are so tightly regulated by baryonic matter, which only represents $\sim$10\% of the galaxy mass. Moreover, these pieces of evidence are not intuitively explained in a Universe where the formation of structures occurs hierarchically through a stochastic merging process dominated by~DM.

An even more noteworthy coincidence, which might be difficult to interpret in $\Lambda$CDM, is the emergence of an acceleration scale, $a_0 \simeq 1.2 \times 10^{-10}$~m~s$^{-2}$, from~the three relations. This acceleration scale, expressed in natural units, can also be written as $a_0\sim H_0\sim{\Lambda}^{1/2}$, which is even more puzzling since it seems to suggest that DM and DE phenomenologies are regulated by the same acceleration scale~\cite{Famaey_and_McGaugh_2012}. This is even less intuitive in $\Lambda$CDM.

At last, another issue on the galaxy scale is provided by the complex dynamics of some categories of systems. The~relative contributions of the stellar disk, made of baryonic matter, and~of the DM halo to the overall rotation curves of disk galaxies are degenerate to each other, a~problem known as the ``disk-halo conspiracy''~\cite{vanAlbada_1985,vanAlbada_and_Sancisi_1986}. Many pieces of evidence suggest the ``maximum-disk hypothesis''~\cite{vanAlbada_and_Sancisi_1986,Sackett_1997,Courteau_and_Rix_1999,Bissantz_and_Gerhard_2002,Sellwood_and_Debattista_2014,McGaugh_and_Schombert_2015}, where the disk maximally contributes to the inner part of the rotation curve. However, this assumption is only valid for high surface brightness (HSB) galaxies, whose rotation curves are steeply rising in their central regions and are described with a cuspy DM density profile. Instead, dwarf, low surface brightness (LSB), and~dwarf spheroidal (dSph) galaxies are DM-dominated even in their innermost regions and, thus, their rotation curves, slowly rising in their central part and modelled with a cored DM density profile, cannot be described by a maximum-disk model~\cite{Strigari_2008,DiPaolo_2019b}. These categories of galaxies represent one of the best pieces of evidence of the mass-discrepancy problem on the scale of galaxies since they are among the darkest galaxies in the Universe,\linebreak ~e.g., \cite[]{Mateo_1998}, and~they give rise to the cusp/core problem. Intriguingly, there are systems with similar baryonic masses but completely different dynamics. They are globular clusters (GCs), which seem to be nearly DM-free~\cite{Baumgardt_2005,Jordi_2009,Baumgardt_2009,Sollima_and_Nipoti_2010,Ibata_2011a,Ibata_2011b,Frank_2012}.

Some of these problems are more intuitively explained, rather than with the $\Lambda$CDM paradigm, with~a modification of the law of gravity with respect to standard gravity, without~introducing any dark component. According to the theory modified Newtonian dynamics (MOND)~\cite{Milgrom_1983a,Milgrom_1983b,Milgrom_1983c}, the~law of gravity departs from the Newtonian one when the acceleration goes below $a_0$. This theory not only reproduces but actually predicted some pieces of evidence on the galaxy scale, such as the three mentioned scaling relations and the difference between the dynamics of HSB and LSB~galaxies. 

Another theory of modified gravity, more recently introduced, is refracted gravity (RG)~\cite{Matsakos_and_Diaferio_2016}, a~classical theory where the modification of the law of gravity is regulated by the value of the local mass density, rather than of the acceleration. RG has shown some encouraging results in describing the dynamics of galaxies of different shapes, such as disk~\cite{Cesare_2020b} and elliptical E0 galaxies with nearly spherical morphology~\cite{Cesare_2022}. In~RG, the~shape of the gravitational field lines depends on the morphology of the system: they are refracted towards the mid-plane of flattened systems and they remain radial for spherical systems, which might intuitively explain the different dynamics of dwarf galaxies (more flattened) and globular clusters (nearly spherical). Moreover, the~relativistic formulation of RG~\cite{Sanna_2021} belongs to the class of scalar-tensor theories and introduces a single scalar field that explains the phenomenology of both DM and DE. The~covariant RG provides a natural explanation for the $a_0\sim\Lambda^{1/2}$ relation, suggesting a unification of the two dark~sectors.

In this review, we present more intuitive explanations, with~the theories MOND and RG, for~the three scaling relations, for~the emergence of the same acceleration scale from these relations and from the DE sector, and~for the different dynamics of LSB, dwarf, and~dSph galaxies and of GCs. The~outline of the paper develops as follows. In~Section~\ref{sec:Baryonic_scaling_relations}, we detail the BTFR, the~MDAR, and~the RAR (Section~\ref{sec:BTFR_MDAR_RAR}) and we explain the different interpretations of these relations in a $\Lambda$CDM framework (Section~\ref{sec:Newton_BTFR_MDAR_RAR}). Section~\ref{sec:MOND_RG} describes the formulation of MOND and RG theories (Section~\ref{sec:MOND_RG_formulation}) and the interpretation of the three scaling relations with these theories (Section~\ref{sec:MOND_RG_BTFR_MDAR_RAR}). Section~\ref{sec:a0} illustrates how the $a_0$ acceleration scale and the $a_0$--$\Lambda$ relation emerge in Newtonian, MOND, and~RG theories. Section~\ref{sec:Dwarfs_GCs} describes the different dynamical properties of LSB, dwarf, and~dSph galaxies and of GCs and how they can be modelled or interpreted in Newtonian, MOND, and~RG gravities. Section~\ref{sec:Conclusions} concludes the~paper.

\section{The Baryonic Scaling~Relations}
\label{sec:Baryonic_scaling_relations}

The mass discrepancy on the galaxy scale can be neatly quantified by three relations that tightly correlate the properties of dark and baryonic matters in galaxies: the BTFR, the~MDAR, and~the~RAR.

\subsection{Description of the Three~Relations}
\label{sec:BTFR_MDAR_RAR}

\begin{figure}[H]
  
    		\includegraphics[scale=1.80]%
    		{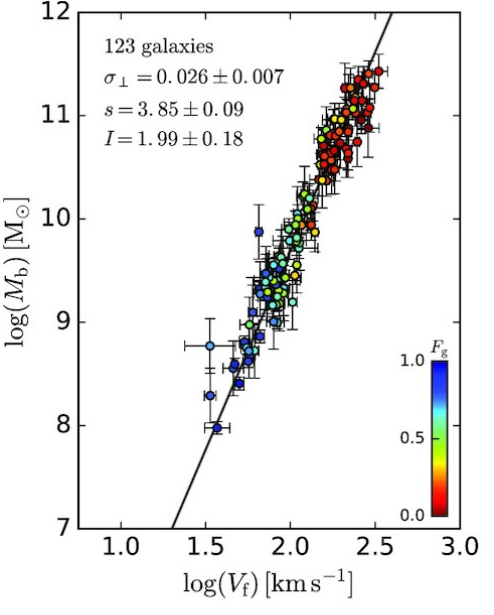}
    		\caption{The baryonic Tully–Fisher relation (BTFR) built for a subsample of 123 galaxies from the SPARC catalogue~\cite{Lelli_2016b}. $M_{\rm b}$ is the total galaxy baryonic mass, accounting for the stars and the gas, and~$V_{\rm f}$ is the mean circular velocity along the flat region of the rotation curve. The~black solid line represents the linear fit to the data with the relation $\log(M_{\rm b}/{\rm M}_\odot) = s\log(V_{\rm f}/\text{km~s}^{-1})+I$. $\sigma_\perp$ denotes the intrinsic scatter of the BTFR, expressed in dex. The~galaxies are colour-coded according to the gas fraction $F_{\rm g}$ to the total baryonic mass. The~figure is re-adapted from Figure~2 in~\cite{Lelli_2019}.}
    		\label{fig:BTFR}

\end{figure}

The BTFR~\cite{McGaugh_BTFR_2000} (Figure~\ref{fig:BTFR}) correlates the total baryonic mass, $M_{\rm bar}$, and~the asymptotic value of the flat part of the rotation curve of galaxies, $V_{\rm f}$, 
according to the relation,~e.g., \cite[]{McGaugh_2012}:
\begin{equation}
\label{eq:BTFR}
    M_{\rm bar} = A V_{\rm f}^b,
\end{equation}
where the normalisation $A$ and the slope $b$ are free parameters. In~the BTFR, each point corresponds to one galaxy. McGaugh~\cite{McGaugh_2012} fits the ($V_{\rm f}$,$M_{\rm bar}$) data points from 47 gas-rich galaxies with Equation~\eqref{eq:BTFR}, adopting different techniques. The~normalisations resulting from these fits are in agreement with each other and the slopes are consistent with 4. In~the BTFR, the~acceleration scale $a_0$ emerges from its normalisation. Setting the slope to 4, McGaugh~\cite{McGaugh_2012} found the normalisation $A = (47 \pm 6)$~M$_\odot$~km$^{-4}$~s$^4$, comparable to the expression $(Ga_0)^{-1}$. For~mass-to-light ratios in the 3.6~$\upmu$m band $M/L_{[3.6]} \gtrsim 0.5$~M$_\odot$/L$_\odot$, the~BTFR intrinsic scatter is minimised to $\sim$0.10~dex~\cite{Lelli_2016a}. The~residuals of the measured BTFR from the model (Equation~\eqref{eq:BTFR}) do not correlate with galaxy properties, such as the radius or the surface brightness~\cite{Desmond_and_Wechsler_2015}.

\begin{figure}[H]
    
    		\includegraphics[scale=0.77]%
    		{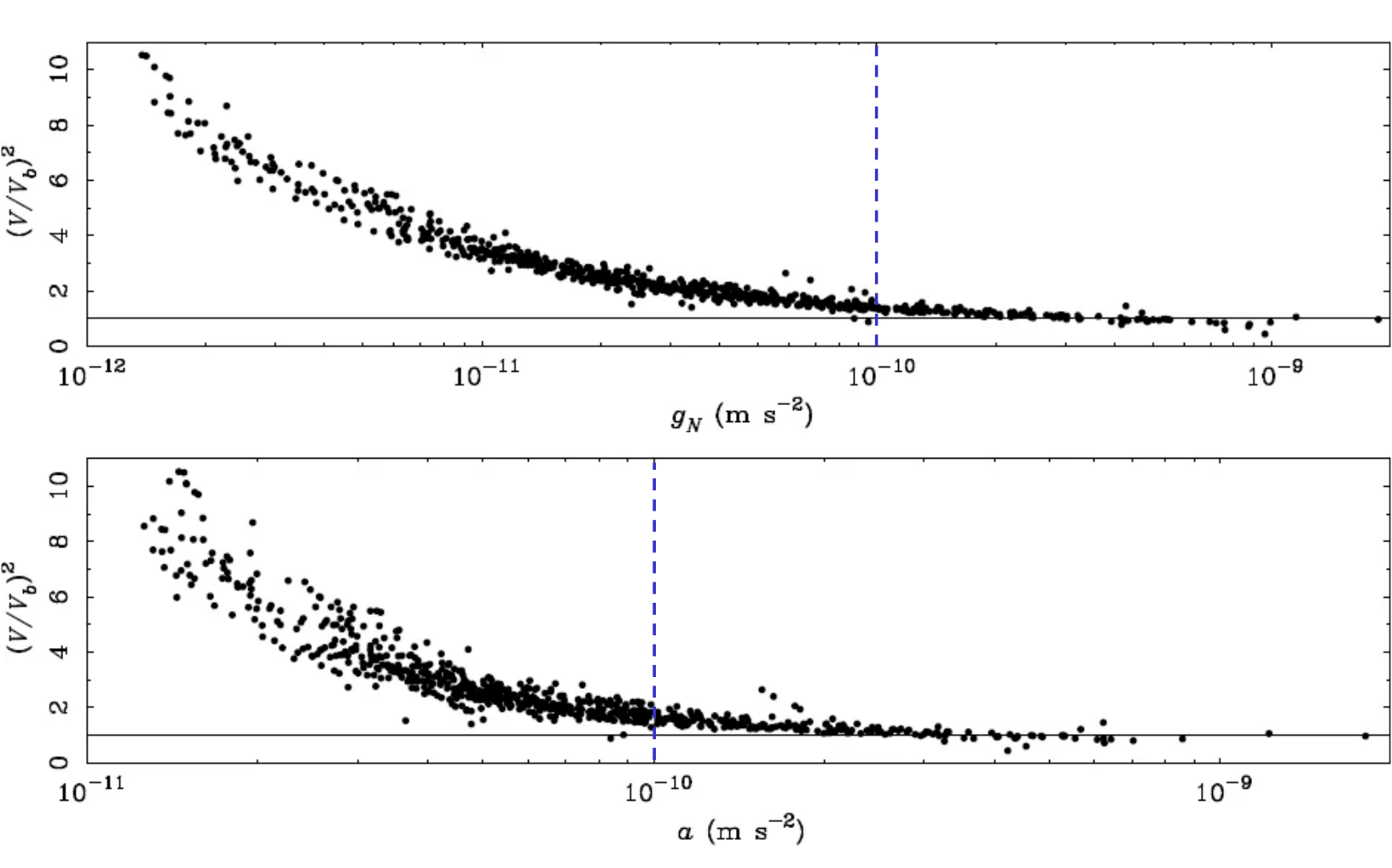}
    		\caption{The mass discrepancy–acceleration
relation (MDAR) represented as the squared ratio between the total ($V$) and the baryonic ($V_{\rm b}$) velocities as a function of the Newtonian acceleration due to baryonic matter, $g_N$ (\textbf{top panel}) and the total acceleration $a$ (\textbf{bottom panel}). We can clearly see that the mass discrepancy is observed from accelerations smaller than $a_0 \simeq 10^{-10}$~m~s$^{-2}$ (indicated in the figure as a vertical blue dashed line, for~reference). The~$(V/V_{\rm b})^2-a$ relation shows a slightly larger scatter than the $(V/V_{\rm b})^2-g_N$ relation. The~black dots represent hundreds of individual resolved data points belonging to the rotation curves of about one hundred spiral galaxies. The~black solid line $(V/V_{\rm b})^2=1$, showing the no mass discrepancy case, is illustrated as a reference in both panels. The~figure is re-adapted from Figure~10 in~\cite{Famaey_and_McGaugh_2012}.}
    		\label{fig:MDAR}

\end{figure}

Two local, rather than global, scaling relations related to the BTFR are the MDAR and the RAR~\cite{Lelli_2017_RAR}. The~MDAR~\cite{McGaugh_MDAR_2004} anti-correlates, at~each distance $R$ from the galaxy centre, the~Newtonian acceleration generated by the baryons distribution, $g_{\rm bar}$, and~the squared ratio $(V/V_{\rm bar})^2$, where $V$ and $V_{\rm bar}$ are the total and the baryons-only velocities (Figure~\ref{fig:MDAR}, top panel). If~we assume spherical symmetry, the~ratio $(V/V_{\rm bar})^2$ coincides with the mass discrepancy, $M/M_{\rm bar}$, where $M$ and $M_{\rm bar}$ are the masses of the entire galaxy and of its baryonic component. The~MDAR can be modelled by the following relation,~e.g., \cite[]{Mayer_2023}:
\begin{equation}
    \label{eq:MDAR}
    \frac{M}{M_{\rm bar}}(R) = 1 + \frac{a_0}{g_{\rm bar}(R)}.
\end{equation}

As long as the acceleration $g_{\rm bar}$ is $\gtrsim a_0$, the~mass discrepancy maintains around 1. Instead, when $g_{\rm bar}$ goes below $a_0$, the~mass discrepancy starts to increase. The~intrinsic scatter in both the MDAR and the BTFR is minimised by the same mass-to-light ratio, consistent with the estimates from stellar population synthesis (SPS) models~\cite{McGaugh_MDAR_2004}. In~fact, the~MDAR can also be expressed as a relation between the mass discrepancy and $g_{\rm obs}$ (Figure~\ref{fig:MDAR}, bottom panel), that is, the~total observed acceleration, which presents a slightly larger scatter than the relation with $g_{\rm bar}$~\cite{Famaey_and_McGaugh_2012}. This relation can be reproduced by replacing $g_{\rm bar}$ with $g_{\rm obs}$ in Equation~\eqref{eq:MDAR}.

\begin{figure}[H]
  
    		\includegraphics[scale=0.85]%
    		{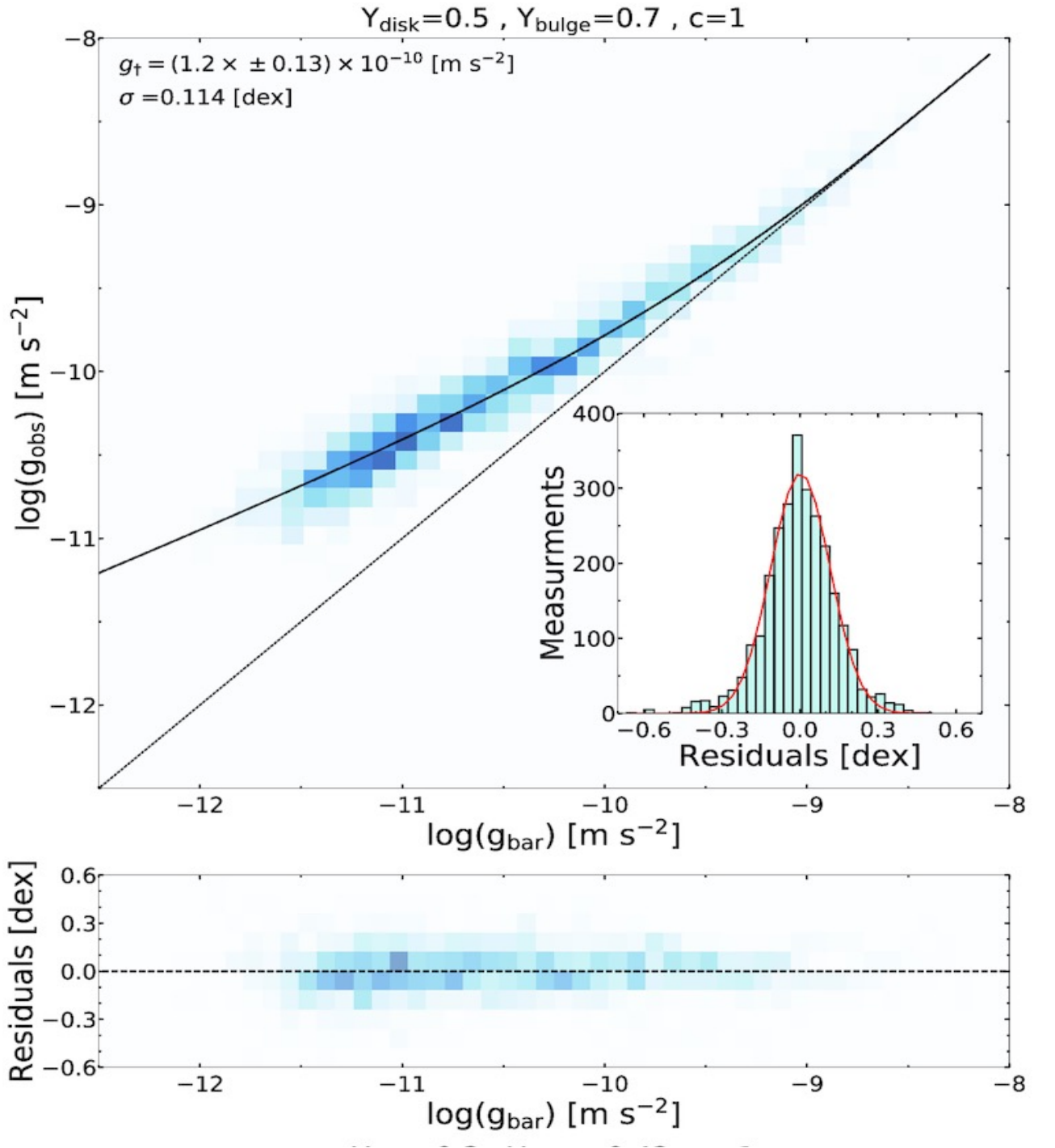}
    		\caption{The radial acceleration relation (RAR) built from Spitzer photometry and accurate rotation curves (SPARC) galaxies. For~each galaxy, a~mass-to-light ratio in the 3.6~$\upmu$m band for the disk and the bulge equal to 0.5 and 0.7~M$_\odot$/L$_\odot$, respectively, is adopted. The~black solid line represents the fit to the data with Equation~\eqref{eq:RAR} and the black dashed line is the $g_{\rm obs} = g_{\rm bar}$ relation, for~reference. The~only best-fit parameter, $g_\dagger$, is highlighted in the top-left corner of the panel and it is 1$\sigma$ consistent with the acceleration scale $a_0$. The~histogram and the bottom panel show the distribution of the residuals of the observed RAR from Equation~\eqref{eq:RAR}, and~its standard deviation $\sigma$, which quantifies the RAR observed scatter, is shown in the top-left corner of the panel. The~figure is re-adapted from Figure~2 in~\cite{Ghari_2019}.}
    		\label{fig:RAR}

\end{figure}

A slightly different perspective with respect to the MDAR is provided by the RAR~\cite{Milgrom_2016} (Figure~\ref{fig:RAR}), which links the centripetal acceleration inferred from the observed rotation curve $V(R)$, $g_{\rm obs} = V^2/R$, with~the Newtonian acceleration due to the baryons alone, $g_{\rm bar}$~\cite{McGaugh_RAR_2016}. The~data in the ($g_{\rm bar}$,$g_{\rm obs}$) plane of the RAR present an even tighter relation with respect to the data in the ($g_{\rm bar}$,$(V/V_{\rm bar})^2$) plane of the MDAR. Moreover, the~($g_{\rm bar}$,$g_{\rm obs}$) plane presents an advantage compared to the ($g_{\rm bar}$,$(V/V_{\rm bar})^2$) plane, since the $g_{\rm bar}$ and $g_{\rm obs}$ quantities and their corresponding uncertainties are completely independent from each other~\cite{Li_2018}.

McGaugh and coauthors~\cite{McGaugh_RAR_2016} fitted the ($g_{\rm bar}$,$g_{\rm obs}$) data from 153 edge-on disk galaxies belonging to Spitzer photometry and accurate rotation curves (SPARC) catalogue~\cite{Lelli_2016b} with the relation:
\begin{equation}
    \label{eq:RAR}
    g_{\rm obs}(R) = \frac{g_{\rm bar}(R)}{1 - \exp\left(-\sqrt{\frac{g_{\rm bar}(R)}{g_\dagger}}\right)},
\end{equation}
where the only free parameter $g_\dagger = (1.20 \pm 0.02 \pm 0.24) \times 10^{-10}$~m~s$^{-2}$ is 1$\sigma$ consistent with $a_0$. The~errors of $0.02 \times 10^{-10}$~m~s$^{-2}$ and of $0.24 \times 10^{-10}$~m~s$^{-2}$ represent the random and the systematic contributions to the uncertainty, respectively. In~particular, the~random error represents the 1$\sigma$ confidence interval and the systematic error represents the 20\% normalisation uncertainty due to the fact that the mass-to-light ratios of the disks and the bulges of the galaxies are kept fixed across the SPARC~sample.

Assuming a disk and a bulge mass-to-light ratio $M/L_{[3.6]}$ of 0.5 and 0.7~M$_\odot$/L$_\odot$, respectively, which are reasonable values in the 3.6~$\upmu$m band, the~RAR of SPARC galaxies is retrieved with an observed scatter of 0.13~dex~\cite{McGaugh_RAR_2016}. This value closely coincides with the scatter of 0.12~dex due to the observational errors of the measured rotation curves, distances, and~galaxy inclinations, and~to the possible variation of the mass-to-light ratio among galaxies, which leaves little room for intrinsic scatter~\cite{McGaugh_RAR_2016,Lelli_2017_RAR}. 

Li and collaborators~\cite{Li_2018} wanted to test whether the RAR was followed by the individual galaxies in the SPARC sample. They fitted Equation~\eqref{eq:RAR} to the observed RAR of the individual SPARC galaxies, finding a RAR with a smaller scatter of [0.054--0.057]~dex, both by fixing $g_\dagger$ to $a_0$ and by leaving it free to vary. Since, differently from~\cite{McGaugh_RAR_2016}, they estimated the mass-to-light ratios and marginalised over the errors on the galaxy distances and inclinations from the RAR of the single SPARC galaxies, the~obtained scatter might be assimilated to the intrinsic one, rather than to the observed~one.

The three relations cover a baryonic mass range of six orders of magnitude, from~$M_{\rm bar}\sim10^{12}$~$M_\odot$, correspondent to the most massive HSB spiral galaxies, to~$M_{\rm bar}\sim10^6$~$M_\odot$, correspondent to the dwarf and the LSB galaxies. However, the~observed scatter of the RAR might increase from 0.13~dex to 0.24~dex for small $g_{\rm bar}$ due to dwarf and LSB galaxies with slowly-rising rotation curves~\cite{Santos-Santos_2020}, which requires further~investigation.

The works of McGaugh~et~al. and of Kroupa~et~al.~\cite{McGaugh_2018,Kroupa_2018} might demonstrate a fundamental origin for the acceleration scale $a_0$, and, thus, for~the RAR, since they show a consistency of $g_\dagger$ among different galaxies. The~work of Li~et~al.~\cite{Li_2018} might suggest the same result, since fitting Equation~\eqref{eq:RAR} to the RAR of the individual SPARC galaxies by leaving $g_\dagger$ free to vary does neither improve the $\chi^2$ nor reduce the obtained scatter with respect to the fits in which $g_\dagger$ is fixed to $a_0 = 1.2 \times 10^{-10}$~m~s$^{-2}$. However, several works question this result. For~example, Rodrigues~et~al.~\cite{Rodrigues_2018a} estimated the $g_\dagger$ parameter with Bayesian inference from 193 individual galaxies from SPARC~\cite{Lelli_2016b} and THINGS~\cite{Walter_2008,deBlok_2008} samples, rejecting the consistency of $g_\dagger$ among the galaxies at the 10$\sigma$ level. The~discrepancy reduces but does not disappear by removing some approximations, becoming equal to 5$\sigma$~\cite{Marra_2020}. The~work of Zhou~et~al.~\cite{Zhou_2020} leads to a similar result. If~this is the case, it would represent an important challenge for MOND (see Section~\ref{sec:MOND_RG_BTFR_MDAR_RAR}). A~debate about the existence or the absence of a universal $a_0$ is presented in~\cite{Rodrigues_2018b}.
Another study sees the emergence of an acceleration scale consistent with $a_0$, besides~from the BTFR, built from rotationally-supported galaxies, from~the baryonic Faber--Jackson relation built from pressure-supported galaxies, such as elliptical galaxies and GCs, and~from the baryonic Faber--Jackson relation built from galaxy clusters (\cite{Edmonds_2020} and references). The~fact that the same acceleration scale also emerges from the dynamics of galaxy clusters, which are different from galaxies both in terms of scale and evolutionary histories, might pose an even more severe issue \mbox{for $\Lambda$CDM.}

\subsection{Interpretation of the Three Relations in Newtonian~Gravity}
\label{sec:Newton_BTFR_MDAR_RAR}

The $\Lambda$CDM paradigm does not provide a natural explanation for these three relations. Indeed, these relations correlate quantities related to the dynamics of a galaxy to its baryonic content, which is counterintuitive in a universe where the dynamics of structures is dominated by DM. The~pieces of evidence that are most difficult to interpret are the small scatter, in~agreement with observational uncertainties, and~the lack of correlations of the residuals between the observed and the modelled relations with the galaxy properties. This phenomenology indicates a quite precise fine-tuning between dark and baryonic matter in galaxies. Moreover, the~MDAR and the RAR pose a more serious issue for $\Lambda$CDM compared to the BTFR due to their local, rather than global, nature.

The three relations resulting from semi-analytical models of galaxy formation and DM-only cosmological simulations in $\Lambda$CDM do not completely agree with the observational data. The~predicted minimum scatter of the BTFR is larger than the observed one (0.17~dex vs 0.10~dex)~\cite{Dutton_2012,DiCintio_and_Lelli_2016} and the simulated BTFR slope $b$ is equal to 3, 8$\sigma$ discrepant from the value $b = 3.98 \pm 0.12$, fitted by McGaugh and coauthors~\cite{McGaugh_BTFR_2000} from the observed BTFR. However, some successful explanations for the three relations remaining within the $\Lambda$CDM paradigm exist, even if with some~issues.

Ludlow~et~al.~\cite{Ludlow_2017} demonstrated that a set of galaxies resulting from the EAGLE suite of $\Lambda$CDM hydrodynamical simulations, run with the same initial conditions but with different stellar and active galactic nuclei (AGN) feedback processes for the baryonic component, follow a RAR-like relation, independently from the considered feedback. Different feedback models make simulated galaxies moving along the RAR and not perpendicular to it, which produces a small RAR scatter $\gtrsim 0.08$~dex. However, the~$g_\dagger$ parameter fitted with Equation~\eqref{eq:RAR} from the galaxies simulated in~\cite{Ludlow_2017} results in a 70$\sigma$ inconsistency with \mbox{$a_0 = 1.2 \times 10^{-10}$~m~s$^{-2}$.} Moreover, measurement errors are not included in the simulated galaxies and the obtained result about the scatter should be further investigated~\cite{Li_2018}. The~same consideration can be drawn for the scatter of 0.06~dex of the RAR obtained from the 32 galaxies resulting from the MUGS2 ``zoom-in'' hydrodynamic simulations in $\Lambda$CDM~\cite{Keller_and_Wadsley_2017}. On~the other hand, Stone and Courteau~\cite{Stone_and_Courteau_2019} found that the intrinsic scatter of the stellar RAR, where only stellar mass is considered to compute the baryonic acceleration, is of $(0.11 \pm 0.02)$~dex, in~agreement, despite being slightly larger, with~$\Lambda$CDM predictions. They obtained this result from PROBES, a~catalogue made of more than 2500 spiral galaxies taken from six deep imaging and spectroscopic surveys. Yet, neglecting gas masses might affect the obtained result.
In fact, the~question of the RAR raises further complications since for small $g_{\rm bar}$, namely for dwarf and LSB galaxies, its scatter increases~\cite{Santos-Santos_2020} (see \mbox{Section~\ref{sec:BTFR_MDAR_RAR}),} and~the RAR built from some galaxy samples different from SPARC shows some correlations between its residuals and certain galaxy properties~\cite{DiPaolo_2019a,Cesare_2020b}.

Some hydrodynamical simulations, which are simulations that include the presence of baryons, can reproduce the slope of the BTFR~\cite{Ferrero_2017a,Ferrero_and_Abadi_2017b}. Yet, the~small scatter of the BTFR can only be explained with a quite precise balance between star formation efficiency and stellar feedback processes~\cite{McGaugh_2012,Lelli_2016a}.

With the semi-empirical model proposed by Di Cintio and Lelli~\cite{DiCintio_and_Lelli_2016}, the~shape and the scatter of the MDAR are reproduced but this does not simultaneously account for the small scatter of $\sim$0.10~dex of the BTFR, which instead results in 0.17~dex.

Mayer~et~al.~\cite{Mayer_2023} used the \textit{Magneticum} hydrodynamical simulation, which provides a large and representative sample of galaxies covering a large range of masses and a variety of morphologies, from~$z\sim0$ to $z\sim3$, to~see whether the baryonic scaling relations (BTFR, MDAR, and~RAR) were reproduced in a $\Lambda$CDM context. The~resulting BTFR, built for galaxies at different redshifts in the range of $0.1 < z < 2.3$, has a slope more consistent with 3, rather than with 4, as~in the observed BTFR. The~MDAR and the RAR built from simulated \textit{Magneticum} galaxies at redshift $z\sim0.1$ reproduce the observed relations (Equations~\eqref{eq:MDAR} and~\eqref{eq:RAR}) with a fitted acceleration scale consistent with $a_0 = 1.2 \times 10^{-10}$~m~s$^{-2}$ and a scatter in agreement with the observations from the SPARC sample. However, the~simulated MDAR and RAR show a positive correlation between the total baryonic mass and the mass discrepancy which is not observed in SPARC data. Other $\Lambda$CDM hydrodynamical simulations of single galaxies might indicate that the baryonic scaling relations naturally emerge in this framework as well~\cite{Navarro_2017,Dutton_2019}.

It is important to note that many small-scale problems of $\Lambda$CDM, such as too large bulges in disk galaxies, the~cusp/core problem, and~the too-big-to-fail problem, have been, at~least partially, solved by introducing baryonic feedback mechanisms, such as outflows due to AGN and Supernovae, and, therefore, it is not so unlikely to suppose that galaxy dynamics is regulated by baryonic physics~\cite{Mayer_2023}.

\section{MOND and~RG}
\label{sec:MOND_RG}

A totally different scenario for the three scaling relations is provided by theories of modified gravity that do not imply the presence of DM. Without~DM, the~fine-tuning issue between the properties of dark and baryonic matters disappears. A~modified gravity theory that not only describes but even predicted these scaling relations is MOND. Another theory of modified gravity that seems to account for these relations is RG. In~the following subsections, I present the formulation of these two theories of gravity and how they reproduce the mentioned scaling~relations.

\subsection{Summary of Theories~Formulation}
\label{sec:MOND_RG_formulation}

\subsubsection{MOND}
\label{sec:MOND_formulation}

In 1983, Milgrom~\cite{Milgrom_1983a,Milgrom_1983b,Milgrom_1983c} formulated MOND, a~theory of gravity that mimics the effect of DM with a boost of the gravitational field compared to the Newtonian one in low-acceleration environments. MOND is a general paradigm that assumes spacetime scale-invariance when the acceleration is $a \ll a_0$. Specifically, the~acceleration $a$ presents the following asymptotic values:
\begin{equation}
    \label{eq:inertiaMONDasymptotic}
    a \simeq
    \begin{cases}
    g_{\rm N}, & a \gg a_0\\
    \sqrt{g_{\rm N}a_0}, & a \ll a_0,
    \end{cases}
\end{equation}
where $g_{\rm N}$ is the Newtonian~acceleration.

The MOND paradigm is obtained by modifying either gravity,~e.g., \cite[]{Bekenstein_and_Milgrom_1984} or inertia~\cite{Milgrom_1994}, where the modified-inertia version~\cite{Milgrom_1994} was less developed. In~the first nonrelativistic modified gravity version of MOND~\cite{Bekenstein_and_Milgrom_1984}, the~following Poisson equation was defined:
\begin{equation}
    \label{eq:PoissonMOND}
    	\nabla \cdot \left[\mu\left(\frac{\left\vert \nabla\phi \right\vert}{a_0}\right) \nabla\phi \right] = 4\pi G\rho,
\end{equation}
where the interpolating function $\mu$, monotonic in its argument, has these two asymptotic behaviours:
\begin{equation}
    \label{eq:muMOND}
    	\mu\left(\frac{\left\vert \nabla\phi \right\vert}{a_0}\right) =
    	\begin{cases}
    		1, & a \gg a_0\\
    		\frac{\left\vert \nabla\phi \right\vert}{a_0}, & a \ll a_0.
    	\end{cases}
\end{equation}

For $a \gg a_0$, the~Newtonian Poisson equation is retrieved:
\begin{equation}
    \label{eq:PoissonN}
    	\nabla^2\phi_{\rm N} = 4\pi G\rho,
\end{equation}
and for $a \ll a_0$, we observe a boost of the gravitational field over the Newtonian one. In~this regime, whereas the Newtonian field is $\propto R^{-2}$, the~MOND field is $\propto R^{-1}$, deviating from the Newtonian inverse square law and reproducing the flat trend of the rotation curves without the presence of~DM.

The MOND paradigm was also defined with other modified gravity formulations, such as QUMOND~\cite{Milgrom_2010}, where the MONDian behaviour of the gravitational field is obtained with the Poisson equation:
\begin{equation}
    \label{eq:PoissonQUMOND}
    \nabla^2 \phi = \nabla \cdot \left[\nu\left(\frac{\left\vert \nabla\phi_{\rm N}\right\vert}{a_0}\right)\nabla\phi_{\rm N}\right],
\end{equation}
where $\phi_{\rm N}$ is Newtonian gravitational potential. A~possible form of the interpolating function $\nu$ is given by the ``simple $\nu$-function'' (Equation (50) with $n = 1$ in~\cite{Famaey_and_McGaugh_2012}):
\begin{equation}
    \label{eq:nuMOND}
    \nu(y) = \frac{1}{2}\left(1 + \sqrt{1 + \frac{4}{y}}\right),
\end{equation}
with $y$ being equal to $\frac{\left\vert \nabla\phi_{\rm N}\right\vert}{a_0}$.

\subsubsection{Refracted~Gravity}
\label{sec:RG_formulation}

RG is a novel theory of modified gravity inspired by the behaviour of electrodynamics in matter that does not resort to DM~\cite{Matsakos_and_Diaferio_2016}. RG was formulated in a nonrelativistic way by~\cite{Matsakos_and_Diaferio_2016} and its gravitational potential $\phi$ obeys the modified Poisson equation:
\begin{equation}
    \label{eq:PoissonRG}
    	\nabla \cdot \left[\epsilon(\rho) \nabla\phi \right] = 4\pi G\rho,
\end{equation}
where the gravitational permittivity $\epsilon(\rho)$ mimics the DM~phenomenology.

Whereas in MOND the modification of the law of gravity is regulated by an acceleration scale, in~RG it is regulated by a mass density scale. The~gravitational permittivity $\epsilon(\rho)$ is a monotonic increasing function of the local mass density $\rho$, it depends on three universal free parameters, and~it presents the following asymptotic limits in the high and low-density regimes:
\begin{equation}
    \label{eq:epsilonRGasymptreg}
    	\epsilon(\rho) \simeq
    	\begin{cases}
    		1, & \rho \gg \rho_{\rm c}\\
    		\epsilon_0, & \rho \ll \rho_{\rm c},
    	\end{cases}
\end{equation}
where the permittivity of vacuum $\epsilon_0$ and the critical density $\rho_{\rm c}$ are two of the three free parameters of the~theory.

As in MOND for $a \gg a_0$, for~$\rho \gg \rho_{\rm c}$, Equation~\eqref{eq:PoissonRG} reduces to the Newtonian Poisson equation~\eqref{eq:PoissonN}. When $\rho \ll \rho_{\rm c}$, the~RG field is boosted compared to the Newtonian one thanks to the value of $\epsilon_0 \in (0,1)$.

RG predicts a different behaviour for the gravitational field in the low-density environments of spherical and flattened systems. In~the external regions of spherical systems, where $\rho \ll \rho_{\rm c}$, the~gravitational field does not deviate from the inverse square law, i.e.,~$\partial\phi/\partial r \propto r^{-2}$, as~in the Newtonian case, but~it is boosted with respect to the Newtonian field by the inverse of the gravitational permittivity, as~obtained by integrating Equation~\eqref{eq:PoissonRG}:
\begin{equation}
    \label{eq:RGfieldsph}
    \frac{\partial\phi}{\partial r} = \frac{1}{\epsilon(\rho)} \frac{G M(<r)}{r^2} = \frac{1}{\epsilon(\rho)} \frac{\partial\phi_{\rm N}}{\partial r},
\end{equation}
where $M(<r)$ is the system mass within the spherical radius $r$.

Whereas for spherical systems we still observe a Newtonian trend, in~the outskirts of flattened systems the field lines are refracted toward the mid-plane of the object. This can be seen by expanding the left-hand side of Equation~\eqref{eq:PoissonRG}:
\begin{equation}
    \label{eq:PoissonRGexpanded1stLHS}
    \frac{\partial \epsilon}{\partial \rho} \nabla\rho \cdot \nabla\phi + \epsilon(\rho)\nabla^2\phi = 4\pi G \rho,
\end{equation}
where the term ``$\frac{\partial \epsilon}{\partial \rho} \nabla\rho \cdot \nabla\phi$'', different from zero in nonspherical systems, is responsible for the focussing of the field lines. In~nonspherical configurations, we thus observe an analogy with electrodynamics in matter: the gravitational field lines behave like electric field lines when they cross a dielectric medium with a nonuniform permittivity, changing both in direction and in~magnitude.

\begin{figure}[H]
  
    		\includegraphics[scale=0.40]%
    		{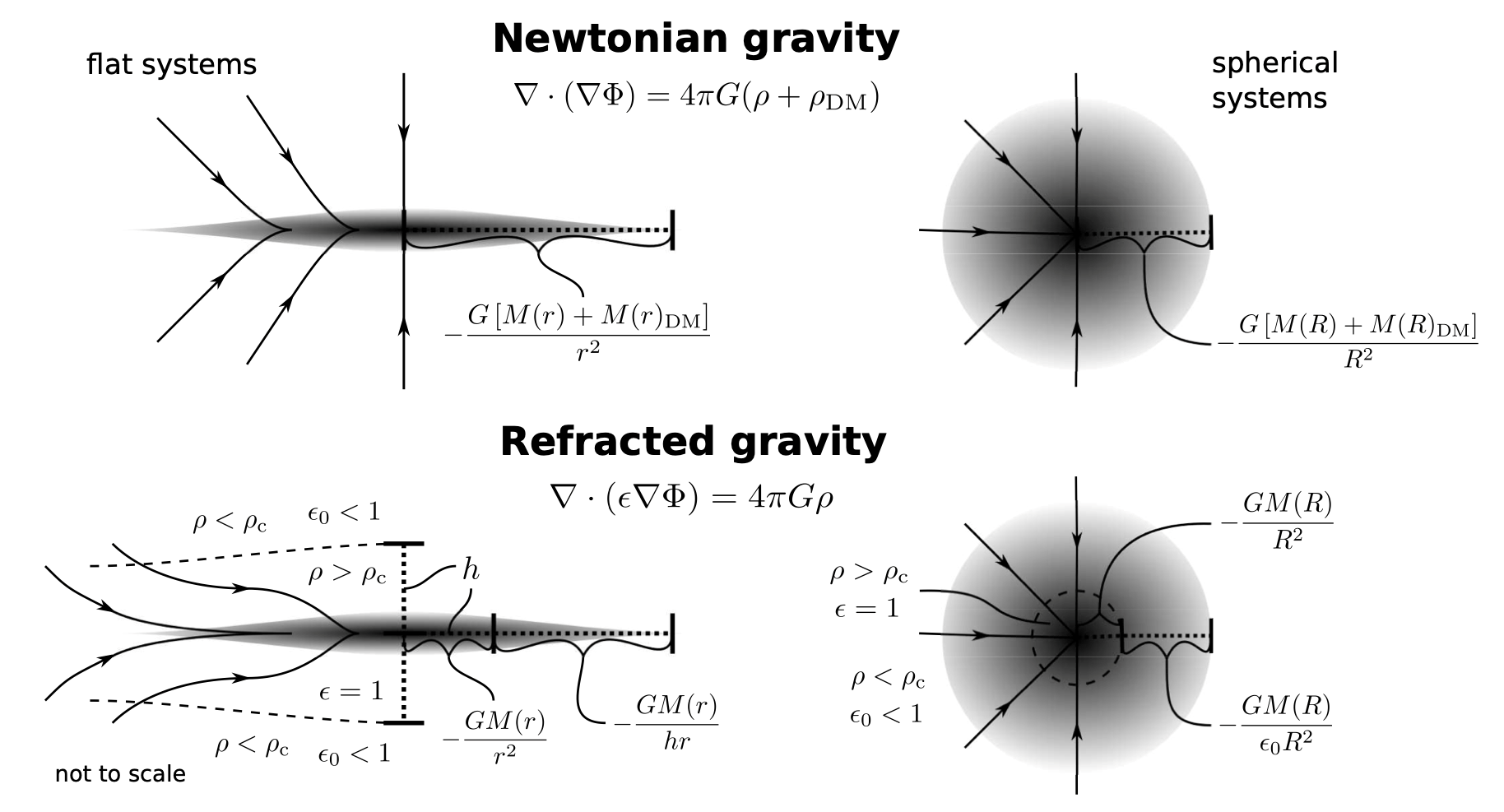}
    		\caption{Comparison between the behaviour of Newtonian (\textbf{top panels}) and refracted gravity (RG) (\textbf{bottom panels}) gravitational fields for flat (\textbf{left panels}) and spherical (\textbf{right panels}) systems. The~figure is taken from Figure~16 in~\cite{Matsakos_and_Diaferio_2016}.}
    		\label{fig:Newton_RG_Flat_Spherical_Systems}
    
\end{figure}

This redirection effect in the low-density regions of flattened systems yields the trend $\partial\phi/\partial R\sim\sqrt{a_0 \vert g_{\rm N} \vert} \propto R^{-1}$ for the RG gravitational field~\cite{Matsakos_and_Diaferio_2016}, where $\vert g_{\rm N} \vert = \vert \partial \phi_{\rm N}/\partial R\vert$ is the Newtonian field. In~this regime, the~RG field deviates from the Newtonian inverse square law and is subject to a boost that in Newtonian gravity is obtained with the presence of DM. This limit also coincides with the MOND asymptotic behaviour for $a \ll a_0$ (see Equation~\eqref{eq:inertiaMONDasymptotic}) and it suggests that the ability of MOND in describing galaxy dynamics is shared by RG, as~demonstrated in~\cite{Cesare_2020b}. According to Equation~\eqref{eq:PoissonRGexpanded1stLHS}, the~flatter the system, the~larger the boost of the gravitational field, and, thus, the~larger the mass discrepancy, as~interpreted in Newtonian theory. Figure~\ref{fig:Newton_RG_Flat_Spherical_Systems} summarises the analogies and the differences between Newtonian (top panels) and RG (bottom panels) gravitational fields for flat (left panels) and spherical (right panels) systems.
\vspace{-6pt}

The behaviour of the RG gravitational field in the low-density regime is comparable to the behaviour of the MOND gravitational field in the low-acceleration regime only for nonspherical configurations. In~the low-acceleration regions of spherical systems, MOND field is $\propto R^{-1}$, as~in the low-acceleration regions of flattened systems, whereas, in~the low-density regions of spherical systems, the~$R$-dependence of RG field remains Newtonian. Moreover, RG and MOND fields also present a difference for flattened systems. As~we can see in the left panels of Figure~\ref{fig:Newton_RG_Flat_Spherical_Systems}, the~refraction effect of the RG field lines in a flattened object already begins where the local density $\rho$ is still above $\rho_{\rm c}$ and, in~that region, the~RG field already begins to be boosted compared to Newtonian one. Instead, MOND modification with respect to the Newtonian field only appears in regions where $a < a_0$.

In all the analyses of galaxy dynamics performed with RG~\cite{Matsakos_and_Diaferio_2016,Cesare_2020b,Cesare_2022}, the~following smooth step function of the gravitational permittivity was adopted:
\begin{equation}
    \label{eq:epsilon}
    \epsilon(\rho)=\epsilon_0+(1-\epsilon_0)\frac{1}{2}\left\{\tanh\left[\text{ln}\left(\frac{\rho}{\rho_\mathrm{c}}\right)^Q\right]+1\right\},
\end{equation}
where the power index $Q$ is the third free parameter of the theory and it regulates the transition speed between Newtonian and RG regimes (the larger its value the steeper the transition).

Being that the modification of the law of gravity is dependent on a scalar quantity, the~mass density of baryonic matter\endnote{In fact, the~gravitational sources depend on other scalar quantities besides the mass density, such as their total mechanical and thermodynamical energy or their entropy. Yet, these quantities depend in turn on the mass density and, thus, adopting a gravitational permittivity also dependent on these quantities would be likely to produce a phenomenology comparable to the one obtained with the simple dependence on the mass density alone~\cite{Matsakos_and_Diaferio_2016,Cesare_2020b}.}, it was possible to build a covariant formulation of RG~\cite{Sanna_2021} without the challenges encountered in defining a relativistic extension of MOND~\cite{Bekenstein_and_Milgrom_1984,Bekenstein_1988,Sanders_1988,Skordis_2009,Bekenstein_2011}. On~the other hand, a~modification of the law of gravity that is dependent on a density scale might appear not so intuitive, given that the majority of the pieces of evidence on galaxy scale rather see the emergence of an acceleration scale, $a_0$, below~which a departure from Newtonian gravity is observed. However, the~acceleration scale $a_0$ also seems to appear in RG, from~the weak field limit (WFL) of covariant refracted gravity (CRG). This point will be better addressed in Sections~\ref{sec:a0} and~\ref{sec:Conclusions}.

\subsection{Interpretation of the Three Scaling Relations in MOND and~RG}
\label{sec:MOND_RG_BTFR_MDAR_RAR}

The fact that the acceleration scale $a_0$ emerges from the three scaling relations seems to identify MOND as the most natural solution to explain them. MOND not only reproduces but actually predicted the three relations with a zero intrinsic scatter many years before they were observed, besides~other pieces of evidence on the galaxy scale. It is the only theory of gravity that has this peculiar feature. Already in its first formulation of 1983~\cite{Milgrom_1983b}, Milgrom concludes that ``The $V^4_\infty = a_0 G M$ relation should hold exactly'', where $V_\infty$ is the asymptotic flat velocity of the rotation curve, $M$ is the total baryonic mass of the galaxy (the total mass in MOND), and~the mentioned equation coincides with the BTFR \mbox{(Equation~\eqref{eq:BTFR}).} The~acceleration scale is set by its normalisation. Later observations confirmed this prediction (see Section~\ref{sec:BTFR_MDAR_RAR}). In~his first works of 1983, Milgrom calculated in several independent ways the value of $a_0$~\cite{Milgrom_1983b}, which turned out to be consistent with the value observed years after from the scaling relations. Moreover, Milgrom predicted that the BTFR does not depend on the galaxy type or on any other galaxy property~\cite{Milgrom_1983b}, in~agreement with the future~observations.

The condition on the acceleration $a < a_0$, where the departure from Newtonian dynamics begins to be observed, can be translated in a condition on the surface mass density, $\Sigma < \Sigma_0$, where $\Sigma_0 = a_0/G$~\cite{Famaey_and_McGaugh_2012}. This implies a relation between the mass discrepancy in galaxies and the acceleration due to baryons, which is observed to hold from HSB to LSB galaxies with a very narrow intrinsic scatter and no dependency on galaxy properties~\cite{Famaey_and_McGaugh_2012}.

Concerning the RAR, MOND predicted this relation with a null intrinsic scatter only for the modified inertia version of MOND and for circular orbits~\cite{Milgrom_1994}. For~the modified gravity versions of MOND, the~RAR is recovered with zero intrinsic scatter only for spherical systems, whereas, for~other systems morphologies, it is retrieved with a very small intrinsic scatter, in~agreement with the observations~\cite{Bekenstein_and_Milgrom_1984,Lelli_2017_RAR,Brada_and_Milgrom_1995}.

A work of Eriksen~et~al.~\cite{Eriksen_2019} seems to challenge MOND in modelling the RAR. Specifically, MOND might present a ``cusp/core like'' issue, different from the classical cusp/core problem of $\Lambda$CDM. This issue is particularly relevant for the modified inertia version of MOND. Moreover, by~fitting the RAR relation (Equation~\eqref{eq:RAR}), in~both modified gravity and modified inertia versions of MOND, from~the observational data of SPARC galaxies, the~best fit acceleration scale $g_\dagger$ might result inconsistent among different~galaxies.

Matsakos and Diaferio~\cite{Matsakos_and_Diaferio_2016} demonstrated that RG reproduces both the BTFR and the MDAR. In~Section~\ref{sec:RG_formulation}, I showed that the asymptotic limit for the gravitational field in low-density environments of nonspherical systems is $\partial\phi/\partial R\sim\sqrt{a_0 \vert g_{\rm N} \vert}$, as~in MOND, in~the low-acceleration regime. In~fact, Matsakos and Diaferio~\cite{Matsakos_and_Diaferio_2016} write this asymptotic limit as:
\begin{equation}
    \label{eq:grav_field_asympt_RG}
    \frac{\partial\phi}{\partial R}\sim\sqrt{b \vert g_{\rm N} \vert},
\end{equation}
where $b$ is an acceleration scale that can be expressed as:
\begin{equation}
    \label{eq:b_RG}
    b = \frac{Gm}{h^2},
\end{equation}
and $\pm h$ is the height from the disk plane where the condition $\rho = \rho_{\rm c}$ is reached. In~a simplified formulation for RG (SRG), whose conclusions can be extended to a more generic RG framework, the~volume within the disk planes $z = -h$ and $z = +h$ is where the redirection effect of the field lines~occurs.

\textls[-15]{In the low-density regime, the~condition on the gravitational field given by \mbox{Equation~\eqref{eq:grav_field_asympt_RG}} can be translated in a condition on the rotation velocity:}
\begin{equation}
    \label{eq:rot_curve_asympt_RG}
    V_{\rm f} = \sqrt{R\frac{\partial\phi}{\partial R}}\sim\sqrt[4]{b G M_{\rm bar}},
\end{equation}
being $\vert g_{\rm N} \vert = G M_{\rm bar}/R^2$. Inverting the above equation,
\begin{equation}
    \label{eq:BTFR_RG}
    M_{\rm bar} = (Gb)^{-1} V_{\rm f}^4,
\end{equation}
we obtain the BTFR (Equation~\eqref{eq:BTFR}) with the correct slope. To~make the normalisation of Equation~\eqref{eq:BTFR_RG} in agreement with the normalisation of the observed BTFR, the~acceleration $b$ has to coincide with $a_0$, which sets a condition on the $z = \pm h$ planes that depends on the galaxy baryonic~mass.

\begin{figure}[H]
  
    		\includegraphics[scale=0.60]%
    		{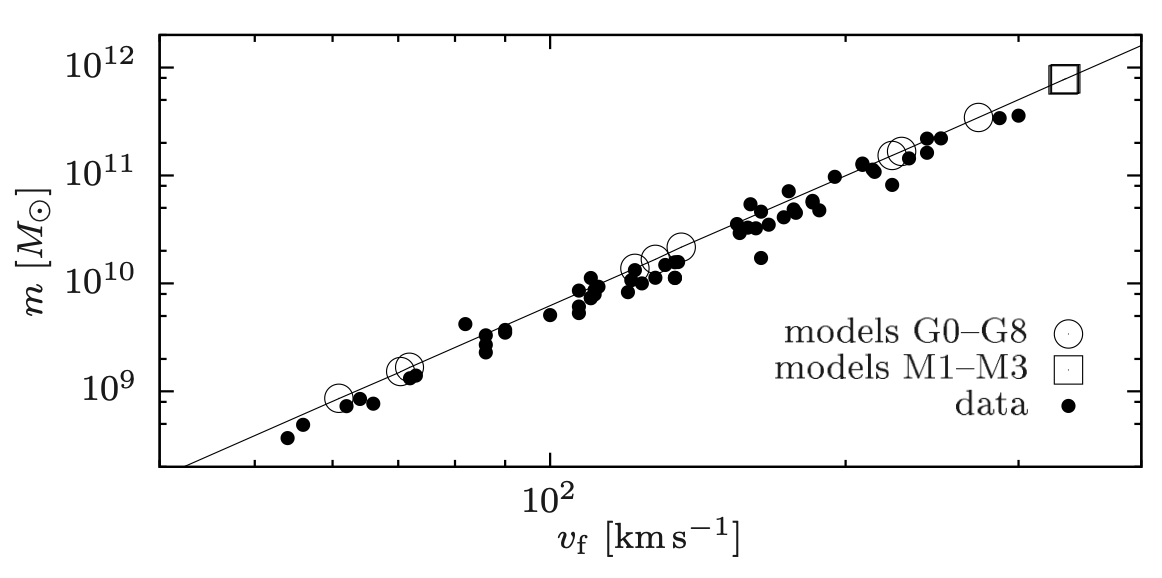}
    		\caption{The BTFR built in RG using Equation~\eqref{eq:BTFR_RG} (black solid line) and from the observational data of~\cite{McGaugh_2005} (black dots) and some models of star- and gas-dominated disk galaxies (open circles and squares). The~RG BTFR properly describes both the data and the model points. The~figure is re-adapted from Figure~9 in~\cite{Matsakos_and_Diaferio_2016}.}
    		\label{fig:BTFR_RG}
  
\end{figure}

Matsakos and Diaferio~\cite{Matsakos_and_Diaferio_2016} plotted the $(V_{\rm f}, M_{\rm bar})$ points from real data~\cite{McGaugh_2005} (black dots in Figure~\ref{fig:BTFR_RG}) and from some disk galaxy models of star- and gas-dominated disk galaxies, with~typical values of the central surface mass density, $\sigma_0$, and~of the disk scale-length, $h_R$ (open circles and squares in Figure~\ref{fig:BTFR_RG}). Both sets of points follow relation~\eqref{eq:BTFR_RG} (black solid line in Figure~\ref{fig:BTFR_RG}).
\vspace{-6pt}

\begin{figure}[H]
  
    		\includegraphics[scale=0.65]%
    		{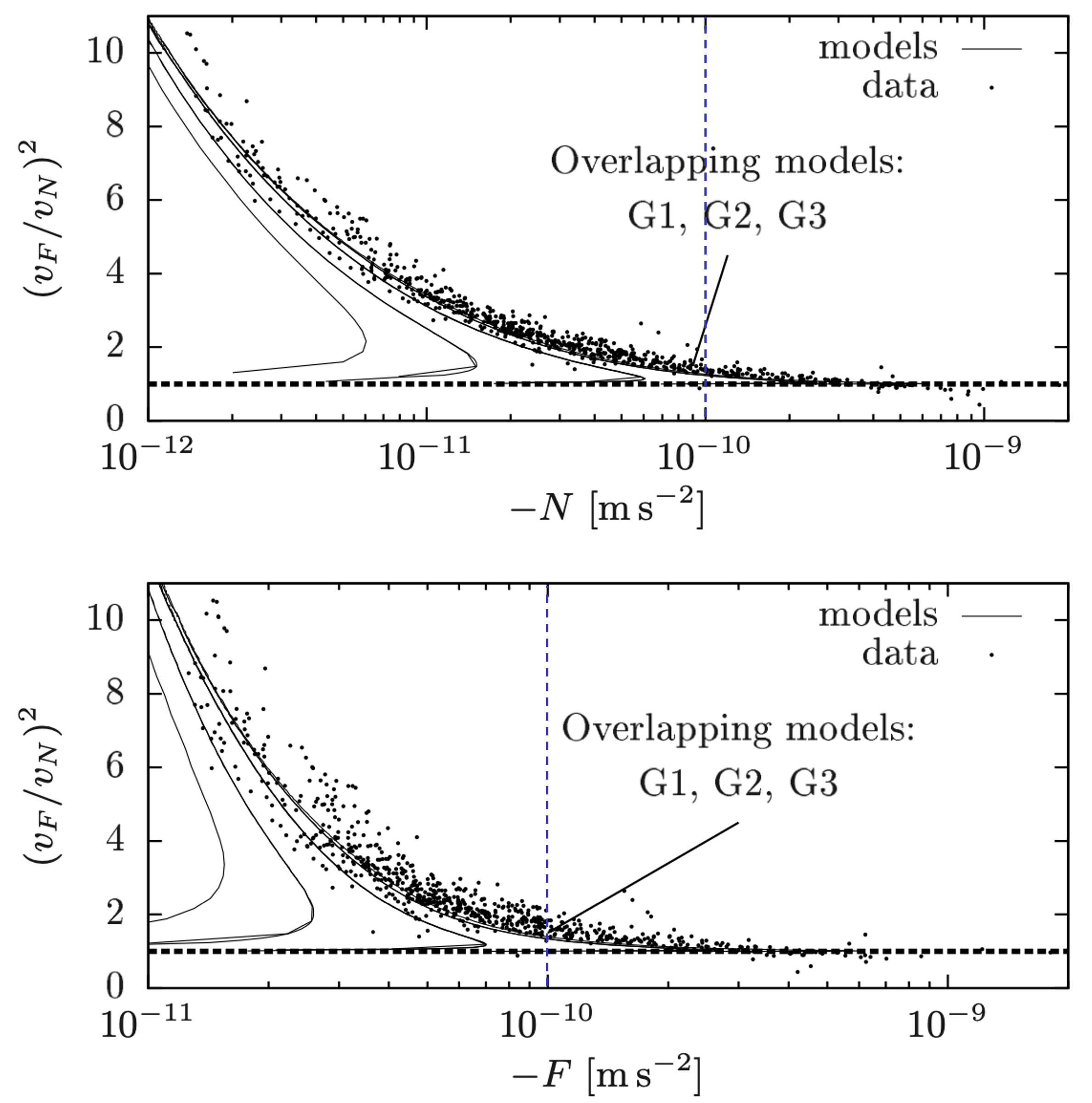}
    		\caption{The MDAR built in RG as the squared ratio between the SRG and the Newtonian speeds, $v_F$ and $v_N$, respectively, as~a function of the Newtonian (\textbf{top panel}) and the SRG (\textbf{bottom panel}) accelerations, $-N$ and $-F$, respectively, for~the same set of models represented as open circles and squares in Figure~\ref{fig:BTFR_RG} (black solid lines) and for a set of point masses having $m = 10^x$~M$_\odot$, with~$x = \{7,8,9,10,11,12\}$ (black dotted lines). The~solid and dotted lines appear almost indistinguishable in the ($-N$,$(v_F/v_N)^2$) and ($-F$,$(v_F/v_N)^2$) planes. The~gas-rich galaxies models appear in the leftmost part of the plots, indicating that they are always in the ``mass discrepancy regime''. The~curves are plotted together with observational measurements from~\cite{Famaey_and_McGaugh_2012,McGaugh_MDAR_2004} (black dots), showing that they properly reproduce the data. The~black dashed line $(v_F/v_N)^2 = 1$, showing the no mass discrepancy case, and~the blue dashed lines $-N = 10^{-10}$~m~s$^{-2}$ and $-F = 10^{-10}$~m~s$^{-2}$ are illustrated as a reference in both panels. The~figure is re-adapted from Figure~10 in~\cite{Matsakos_and_Diaferio_2016}.}
    		\label{fig:MDAR_RG}
  
\end{figure}

RG also recovers the MDAR of the galaxies (Figure~\ref{fig:MDAR_RG}). Matsakos and Diaferio~\cite{Matsakos_and_Diaferio_2016} plotted the squared ratio between the rotation velocity predicted by the SRG framework and the Newtonian theory ($v_F$ and $v_N$ in Figure~\ref{fig:MDAR_RG}) against the Newtonian and the SRG accelerations ($-N$ and $-F$ in Figure~\ref{fig:MDAR_RG}) for the same models used to build the BTFR and for a set of point masses having $m = 10^x$~M$_\odot$, with~$x = \{7,8,9,10,11,12\}$, settling between two parallel planes with $z = \pm h = \pm \sqrt{Gm/b}$. The~models are plotted together with a set of data points, taken from~\cite{McGaugh_MDAR_2004,Famaey_and_McGaugh_2012}, and~they properly reproduce the data, both in the $(g_{\rm bar}, (V/V_{\rm bar})^2)$ and in the $(g_{\rm SRG},(V/V_{\rm bar})^2)$ planes. The~gas-rich galaxies models (where the gas represents more than 20\% of the total baryonic mass) have a smaller $\sigma_0$ and they distribute in the leftmost part of the MDAR, in~agreement with MOND prediction according to which a larger mass discrepancy is observed when $\Sigma < \Sigma_0$. On~the contrary, the~star-rich galaxies curves distribute on the rightmost part of the~MDAR.

\begin{figure}[H]
   
    		\includegraphics[scale=0.80]%
    		{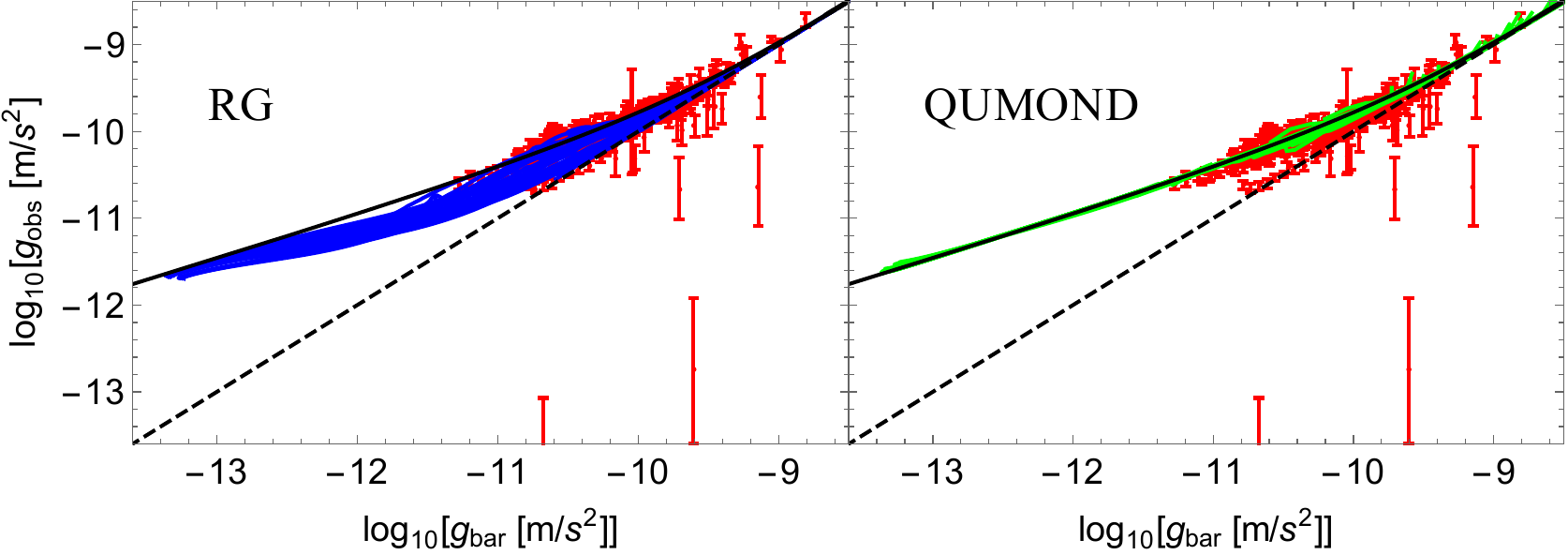}
    		\caption{The RAR built in RG (\textbf{left panel}, blue solid lines) and in QUMOND (\textbf{right panel}, green solid lines) for DMS galaxies. The~RG and QUMOND models are superimposed onto the observed RAR of DMS galaxies (red dots with error bars). Equation~\eqref{eq:RAR} (black solid line) and the $g_{\rm obs} = g_{\rm bar}$ line (black dashed line) are shown as a reference. The~figure is taken from Figure~11 in~\cite{Cesare_2020b}. Credit: Cesare V., Diaferio A., Matsakos T., and~Angus G., A\&A, 637, A70, 2020, reproduced with permission $\copyright$ ESO.}
    		\label{fig:RAR_RG}
  
\end{figure}

Cesare~et~al.~\cite{Cesare_2020b} built the RAR of 30 galaxies from the DiskMass Survey (DMS)~\cite{Bershady_2010a_DMSI}, adopting the general RG framework and the QUMOND formulation of MOND (blue and green solid lines in Figure~\ref{fig:RAR_RG}). They modelled with RG, at~the same time, the~rotation curve and the vertical velocity dispersion profile of each DMS galaxy, obtaining mass-to-light ratios consistent with SPS models, disk scale heights in agreement with the observations of edge-on galaxies, and~RG parameters consistent among the different galaxies, suggesting their universality. To~build the RG acceleration, they numerically solved the RG Poisson equation (Equation~\eqref{eq:PoissonRG}), adopting the mass-to-light ratios, the~disk-scale heights, and~the three RG parameters estimated from the kinematic profiles of each DMS galaxy. To~compute the QUMOND acceleration, they numerically solved the QUMOND Poisson equation (Equation~\eqref{eq:PoissonQUMOND}) with the same mass-to-light ratios and disk scale heights found in RG, since they are fully in agreement with the parameters found by Angus~et~al.~\cite{Angus_2015} by performing the same analysis of the dynamics of DMS galaxies in QUMOND. The~Newtonian acceleration, $g_{\rm bar}$, was calculated by solving the Newtonian Poisson equation (Equation~\eqref{eq:PoissonN}) with the same mass-to-light ratios used to compute the RG and QUMOND accelerations, being in agreement with SPS models, and~with disk-scale heights $h_{z,{\rm SR}}$ derived from the scale relation between the disk-scale lengths and heights:
\begin{equation}
    \label{eq:hR_hz_relation}
    \log_{10} \left(\frac{h_R}{h_{z,{\rm SR}}}\right) = 0.367 \log_{10} \left(\frac{h_R}{\rm kpc}\right) + 0.708 \pm 0.095,
\end{equation}
estimated from 60 edge-on late-type galaxies~\cite{Bershady_2010b_DMSII,Cesare_2020b}.

RG properly reproduces the asymptotic limits of the observed RAR (Equation~\eqref{eq:RAR}, black solid line in Figure~\ref{fig:RAR_RG}) but it tends to underestimate relation~\eqref{eq:RAR} at low $g_{\rm bar}$, even if it generally interpolates the observational data (red dots with error bars in Figure~\ref{fig:RAR_RG}). Instead, QUMOND reproduces the RAR with the correct shape. This can be interpreted by the fact that RG might attribute more luminous mass than~QUMOND.

A more serious problem for RG is that the RAR presents a too large intrinsic scatter (0.11~dex, whereas the possible intrinsic scatter of the RAR found by Li~et~al.~\cite{Li_2018} for SPARC galaxies is 0.057~dex) and some correlations between the residuals from Equation~\eqref{eq:RAR} and certain galaxy properties, in~disagreement with the observations~\cite{McGaugh_RAR_2016,Lelli_2017_RAR}. However, this question has to be further deepened by building the RAR in RG for a larger sample of disk galaxies with more accurate rotation curves, such as SPARC, before~concluding that this is due to an issue of RG theory. Indeed, the~RAR of DMS data also shows some correlations between the residuals and some galaxy properties, which might not be observed in the SPARC sample and suggests that the DMS could not be the most suitable sample where to investigate the~RAR.    

In contrast, the~RAR curves computed in QUMOND, very neatly distribute around Equation~\eqref{eq:RAR} with an intrinsic scatter of 0.017~dex. This is consistent with expectations, since QUMOND is a modified gravity version of MOND, and, thus, it does not provide a RAR with zero intrinsic scatter for nonspherical systems as disk galaxies~\cite{Bekenstein_and_Milgrom_1984}. Moreover, QUMOND RAR also presents correlations between its residuals and some galaxy properties, again in agreement with the fact that, in~this case, the~scatter of the RAR is not equal to zero and which might further suggest that the DMS sample was not the most suitable to investigate the~RAR.

\unskip

\section{Possible Interpretations for an Intriguing Acceleration~Scale}
\label{sec:a0}

If the observed DM-baryons scaling relations might appear not so intuitive in the $\Lambda$CDM context, a~piece of evidence even more difficult to interpret in this framework is the emergence of an acceleration scale, $a_0$, from~the three relations. However, this acceleration scale seems to be retrieved in the $\Lambda$CDM model as well, for~example, from~the MDAR and the RAR built from the simulated galaxies in the \textit{Magneticum} simulation at $z\sim0.1$~\cite{Mayer_2023}. These relations were fitted with Equations~\eqref{eq:MDAR} and~\eqref{eq:RAR}, obtaining an $a_0$ consistent with $1.2 \times 10^{-10}$~m~s$^{-2}$~\cite{Mayer_2023} (see  Section~\ref{sec:Newton_BTFR_MDAR_RAR}). Mayer~et~al.~\cite{Mayer_2023} also found that \textit{Magneticum} galaxies followed a MDAR (Equation~\eqref{eq:MDAR}) and a RAR (Equation~\eqref{eq:RAR}) relation at higher redshifts but with a fitted $a_0$ substantially different. Specifically, $a_0$ decreases for decreasing redshift, that is, in~the more recent Universe. This trend means that the mass discrepancy decreases, and, thus, galaxies become more baryons-dominated, as~cosmic time advances, and~this can be explained by the progressive cooling of the gas during~time.

Mayer~et~al.~\cite{Mayer_2023} tried to assess whether the $a_0(z)$ relation observed in simulated \textit{Magneticum} galaxies is consistent or not with MOND predictions, which is not trivial since MOND is not formulated as a relativistic theory. Milgrom~\cite{Milgrom_1983a} pointed out that $a_0 \thickapprox c H_0$, which might suggest that the redshift dependence of $a_0$ is similar to the one of the Hubble parameter $H(z)$. This would imply the following relation:
\begin{equation}
    \label{eq:Possible_a0z_MOND}
    a_0(z) \thickapprox a_0(0) \times \sqrt{\Omega_{\rm m}(1 + z)^3 + \Omega_\Lambda},
\end{equation}
where the fraction of baryonic matter $\Omega_{\rm m}$ and of the cosmological constant $\Omega_\Lambda$ with respect to the total mass-energy budget of the Universe are put equal to $0.25$ and $0.75$, following~\cite{Bekenstein_and_Sagi_2008}, and~$a_0(0)$ is the value of the acceleration scale $a_0 = 1.2 \times 10^{-10}$~m~s$^{-2}$ observed today. However, this relation is too steep compared to the $a_0(z)$ relation observed in \textit{Magneticum} simulation and it also seems to disagree with observational data. More realistic predictions of the $a_0(z)$ relation in MOND might be provided by relativistic theories that have MOND as a limiting case in a nonrelativistic regime, such as tensor-vector-scalar gravity (TeVeS)~\cite{Bekenstein_TeVeS_2004} and covariant emergent gravity (CEG)~\cite{Hossenfelder_2017}. TeVeS might suggest that $a_0$ varies on timescales larger than the Hubble time, even if, according to~\cite{Bekenstein_and_Sagi_2008}, this change is also possible on cosmological timescales. Anyway, the~$a_0(z)$ relation predicted by TeVeS is not consistent with the one observed in~\cite{Mayer_2023} from \textit{Magneticum} galaxies. CEG predicts an $a_0(z)$ relation dependent on the variation of the size of the cosmological horizon~\cite{Hossenfelder_and_Mistele_CEG_2018}. However, the~redshift dependence of this $a_0(z)$ relation is smaller than the one observed from the galaxies in the MUGS2 simulated sample in $\Lambda$CDM~\cite{Keller_and_Wadsley_2017}.

An even more striking coincidence is given by the near equality between $a_0$ emerging from the local universe, that is, from~galactic dynamics, and~from cosmology. Given the two acceleration cosmological parameters~\cite{Milgrom_a0_cosmology_2020}:
\begin{equation}
    \label{eq:aH}
    a_H \equiv c H_0
\end{equation}
and
\begin{equation}
    \label{eq:aLambda}
    a_\Lambda \equiv c^2 \left(\frac{\Lambda}{3}\right)^{1/2},
\end{equation}
where $H_0$ and $\Lambda$ are the present values of the Hubble parameter and of the cosmological constant, the~following equivalence subsists:
\begin{equation}
    \label{eq:a0_cosmology}
    \bar{a_0} \equiv 2\pi a_0 \thickapprox a_H \thickapprox a_\Lambda.
\end{equation}

This arises an outstanding coincidence that needs careful investigation since $\Lambda$ is assumed to be constant across cosmic time and $H_0$ is the value at the present epoch of the Hubble parameter, $H(t)$, that varies across cosmic~time. 

Some interpretations of the relation $a_0\sim\Lambda^{1/2}$ in the MOND context were provided by~\cite{Famaey_and_McGaugh_2012,Milgrom_1989,Milgrom_1999,Milgrom_a0_cosmology_2020}. In~the most recent work among these, Milgrom~\cite{Milgrom_a0_cosmology_2020} re-wrote Equation~\eqref{eq:a0_cosmology} in terms of length or mass, introducing the ``MOND length'', $l_{\rm M}$, and~the ``MOND mass'', $M_{\rm M}$:
\begin{equation}
    \label{eq:lM}
    l_{\rm M} \equiv \frac{c^2}{a_0} \thickapprox 2\pi l_H \thickapprox 2\pi l_\Lambda
\end{equation}
and
\begin{equation}
    \label{eq:MM}
    M_{\rm M} \equiv \frac{c^4}{G a_0} \thickapprox M_H \thickapprox M_\Lambda,
\end{equation}
where $l_H \equiv cH_0^{-1}$ is the Hubble radius, $l_\Lambda \equiv (\Lambda/3)^{-1/2}$ is the de Sitter radius related to $\Lambda$, and~$M_H$ and $M_\Lambda$ are the total mass of the Universe enclosed within $l_H$ and $l_\Lambda$, respectively.

Equation~\eqref{eq:a0_cosmology} emerges in some particular effective-field, relativistic extensions of MOND, where MOND is retrieved in their WFL. In~these MOND relativistic formulations, an~additional Lagrangian term,
\begin{equation}
    \label{eq:MOND_add_Lagrangian}
    \mathcal{L}_{\rm M} = \frac{c^4}{G} \sqrt{g} \frac{1}{l^2} \mathcal{F}(l^2 Q),
\end{equation}
is added to the GR Einstein--Hilbert (EH) Lagrangian density:
\begin{equation}
    \label{eq:EH_Lagrangian}
    \mathcal{L}_{\rm EH} \propto \frac{c^4}{G} \sqrt{g} R,
\end{equation}
where $R$ is the Riemann curvature scalar. In~Equation~\eqref{eq:MOND_add_Lagrangian}, $l$ is a length constant, needed to provide the Lagrangian with the correct dimension, $\mathcal{F}$ is a dimensionless function, and~$Q$ has the dimension of a length$^{-2}$ and it is built from the first space-time derivatives of the gravitational degrees of freedom~\cite{Milgrom_a0_cosmology_2020}.
Some examples of MOND relativistic extensions where a Lagrangian-like term as Equation~\eqref{eq:MOND_add_Lagrangian} is included are the MOND adaptations of the Einstein--Aether theories~\cite{Zlosnik_2007}, bimetric MOND (BIMOND)~\cite{Milgrom_BIMOND_2009}, bimetric massive gravity~\cite{Blanchet_and_Heisenberg_2017}, and~the noncovariant theory presented in~\cite{Milgrom_noncovariant_MOND_2019}.

If we include the cosmological constant term in the EH Lagrangian, Equation~\eqref{eq:EH_Lagrangian} becomes:
\begin{equation}
    \label{eq:EH_Lagrangian_with_Lambda}
    \mathcal{L}_{\rm EH} \propto \frac{c^4}{G} \sqrt{g} (R - 2\Lambda).
\end{equation}

Therefore, we can see from Equation~\eqref{eq:MOND_add_Lagrangian} that any constant term added to $\mathcal{F}(x)$, which has to be of the order of unity due to naturalness, can be identified with a cosmological constant term of this kind:
\begin{equation}
    \label{eq:Lambda_approx_MOND}
    \Lambda\sim l^{-2} = l_{\rm M}^{-2},
\end{equation}
that, combined with Equation~\eqref{eq:lM}, gives the $a_0$-cosmology coincidence given by\linebreak   \mbox{Equations~\eqref{eq:aLambda} and~\eqref{eq:a0_cosmology}~\cite{Milgrom_a0_cosmology_2020}}. It is important to point out that this result is not obtained by adding a cosmological constant term, $\Lambda$, ``ad-hoc'' to the Lagrangian, verifying a posteriori that $\Lambda\sim(a_0/c^2)^2$, but~by the presence of the two $l$-terms in the Lagrangian~\eqref{eq:MOND_add_Lagrangian}, appearing inside and outside $\mathcal{F}(x)$ for dimensional reasons and playing different roles. This suggests that the two $l$-terms derive from the same underlying physics, whereas a $\Lambda$ constant added ``ad-hoc'' might have suggested a different physical~origin.

Milgrom~\cite{Milgrom_a0_cosmology_2020} also suggested that the $a_0$-cosmology connection can also emerge from a scenario in which the Universe is seen as a sphere-like submanifold, that is a brane, of~radius $l_H$ or $l_\Lambda$, embedded in a space-time with higher dimension~\cite{Milgrom_brane_MOND_2020}. The~dynamics of the submanifold is the one we observe, that is the MOND one, and~it emerges from the dynamics of the higher dimension space-time. In~fact, the~acceleration scale $a_0$ is seen as an emergent acceleration constant in a brane scenario. Further details can be found in~\cite{Milgrom_brane_MOND_2020}.

In RG, where the transition between Newtonian and modified gravity regimes is regulated by a density scale $\rho_{\rm c}$, rather than by an acceleration scale $a_0$, the~$a_0$-cosmology relation might appear harder to interpret. However, Sanna~et~al.~\cite{Sanna_2021} recently formulated a covariant extension of RG, where $a_0$ seems to emerge from the WFL of the theory and the $a_0$-$\Lambda$ coincidence might be~explained.

CRG is a scalar-tensor theory that introduces a single scalar field $\varphi$, nonminimally coupled to the metric, that mimics both the DM effect on galaxy scale and the DE effect on cosmic scale, namely, the~accelerated expansion of the Universe. Therefore, CRG belongs to the restricted class of modified gravity models that invoke a unified dark sector, that is, that attribute the phenomenologies of DM and DE to a single~cause.

CRG is derived from a general scalar-tensor action:
\begin{equation}
    \label{eq:ST_action}
    \mathcal{S} = \frac{1}{16\pi G} \int d^4x \sqrt{g}\left[\varphi R + \frac{\mathcal{W}(\varphi)}{\varphi}\nabla^\alpha\varphi\nabla_\alpha\varphi + 2 \mathcal{V}(\varphi)\right] + \int d^4x \sqrt{g} \mathcal{L}_{\rm m}(g_{\mu\nu},\psi_{\rm m})
\end{equation}
(see~\cite{Sanna_2021} for the explanation of the symbols). In~CRG, the~general differentiable function of the scalar field $\mathcal{W}(\varphi)$ is:
\begin{equation}
    \label{eq:W_varphi_CRG}
    \mathcal{W}(\varphi) = -1,
\end{equation}
and the potential $\mathcal{V}(\varphi)$ has a self-interaction form:
\begin{equation}
    \label{eq:V_varphi_CRG}
    \mathcal{V}(\varphi) = -\Xi\varphi,
\end{equation}
where $\Xi$ is a constant parameter.
With the definitions adopted for $\mathcal{W}(\varphi)$ and $\mathcal{V}(\varphi)$, the~CRG equations below are obtained:
\begin{equation}
    \label{eq:CRG_eq_1}
    \varphi R_{\mu\nu} + \nabla_\mu\nabla_\nu\varphi - \frac{1}{\varphi}\nabla_\mu\varphi\nabla_\nu\varphi = -8\pi GT_{\mu\nu},
\end{equation}
and
\begin{equation}
    \label{eq:CRG_eq_2}
    \square\varphi - 2\Xi\varphi = 8\pi GT.
\end{equation}

CRG is based on a chameleon screening mechanism,~e.g., \cite[]{Khoury_and_Weltman_2004}, that is, in~regions where the Newtonian WFL holds, the~extra degree of freedom of the scalar field $\varphi$ mediates a fifth force which can be detected, whereas, in~high-density regions, this degree of freedom is screened. This behaviour is also what we expect from the RG gravitational permittivity $\epsilon(\rho)$, which, therefore, might be related to $\varphi$.

The WFL of CRG holds the following equation:
\begin{equation}
    \label{eq:WFL_CRG}
    \nabla\cdot (\varphi\nabla\phi) \simeq 8\pi G\rho,
\end{equation}
which reduces to RG Poisson equation~\eqref{eq:PoissonRG} if the scalar field is twice the permittivity, $\varphi = 2\epsilon$. This is an important result since it confirms that the scalar field is associated with the phenomenology mimicked by the gravitational permittivity, that is of DM on galaxy scale. Instead, the~Newtonian Poisson equation is recovered for a constant scalar field $\varphi = 2$. 

Calculating the CRG gravitational field in the WFL from Equation~\eqref{eq:WFL_CRG} for a spherical source with density $\rho_{\rm s}(r)$, monotonically decreasing with $r$, immersed in a homogeneous background with constant density $\rho_{\rm bg}$, Sanna~et~al.~\cite{Sanna_2021} found that an acceleration scale can be set. At~large distances from the source, the~scalar and the gravitational fields, $\varphi$ and $d\phi/dr$, are linked by the relation:
\begin{equation}
    \label{eq:WFL_CRG_sph_source_large_dist}
    \frac{d\ln{\varphi}}{dr} = \frac{d\phi}{dr} \left\{-1 - \left[1 + \left(\frac{d\phi}{dr}\right)^{-2} \left(2\Xi - \frac{8\pi G\rho}{\varphi}\right)\right]^{1/2}\right\},
\end{equation}
where $\rho(r)$ is $\rho_{\rm s}(r) + \rho_{\rm bg}$. The~acceleration scale:
\begin{equation}
     \label{eq:acceleration_scale_CRG}    
     a_\Xi = \left(2\Xi - \frac{8\pi G\rho}{\varphi}\right)^{1/2}
\end{equation}
is set from Equation~\eqref{eq:WFL_CRG_sph_source_large_dist}. In~regions where $d\phi/dr \gg a_\Xi$, the~gravitational field has a similar $r$-dependence as the gravitational field calculated close to the source, that is, the~Newtonian field. Instead, for~$d\phi/dr \ll a_\Xi$, it departs from the Newtonian one. From~this result, the~acceleration $a_\Xi$ recalls $a_0$ since it demarcates Newtonian from modified gravity regimes. Solving the CRG field equations for a homogeneous and isotropic universe with flat curvature, Sanna~et~al.~\cite{Sanna_2021} found that $\Xi\sim\Lambda$, and, thus, $\Xi$ plays the role of the cosmological constant in $\Lambda$CDM. In~the limit obtained at large distances from the source,
\begin{equation}
    \label{eq:CRG_limit_source_large_dist}
    2\Xi \gg 8\pi G\rho/\varphi,
\end{equation}

Equation~\eqref{eq:acceleration_scale_CRG} becomes:
\begin{equation}
    \label{eq:acceleration_scale_CRG_limit_source_large_dist}
    a_\Xi\sim(2\Xi)^{1/2}.
\end{equation}

Using the observed value of $\Lambda$ at the present epoch, Equation~\eqref{eq:acceleration_scale_CRG_limit_source_large_dist} holds $a_\Xi\sim10^{-10}$~m~s$^{-2}$, fully in agreement with the value of $a_0$, and~it also provides the relation $a_0\sim\Lambda^{1/2}$, which is the observed $a_0$-cosmology~connection.

The difference between $a_\Xi$ and $a_0$ consists in the fact that, whereas $a_0$ is a constant independent of the gravitational field source, $\rho$, $a_\Xi$ depends on the source, given Equation~\eqref{eq:acceleration_scale_CRG}, even if for large distances from the source this dependence drops since $\rho_{\rm s} \ll \rho_{\rm bg}$. Future investigations have to verify whether by repeating the calculations for a generic case and not for a specific source, the~connection between $a_\Xi$ and $a_0$ continues to hold in a real~Universe. 

Given the connection between $\Xi$ and the cosmological constant $\Lambda$, it can be concluded that the scalar field $\varphi$ is also responsible for the accelerated expansion of the Universe, that is for the DE phenomenology, besides~the DM phenomenology on galaxy scale, given its connection to the gravitational permittivity $\epsilon$. The~fact that the DM and the DE effects are described by a single scalar field might provide an advantage for CRG since the idea of a unified dark sector is theoretically justified. The~$a_0\sim\Lambda^{1/2}$ relation itself represents one possible piece of evidence for the unification of the dark sector. Moreover, Martin Kunz claims that gravity can only probe the total energy-momentum tensor of the Universe, implying a degeneracy between the dark components, which further goes in this direction~\cite{Kunz_2009a,Kunz_2009b}. 

CRG is not the only theory that provides a single explanation for the DM and DE phenomenologies. Among~the other theories that invoke a unified dark sector we can mention: (I) Conformal gravity~\cite{Mannheim_1989,Nesbet_2013,Campigotto_2019}; (II) the models of Martin Kunz~\cite{Kunz_2009a,Kunz_2009b}; (III) the quartessence model~\cite{Makler_2003,Brandenberger_2019,Ferreira_2019} (known also as Unified Dark Matter or Unified Dark Energy), which includes the generalised Chaplygin gas~\cite{Bento_2002,Carturan_2003,Sandvik_2004}, the~\mbox{$K$-essence~\cite{Scherrer_2004,Giannakis_and_Hu_2005},} the~fast transition models~\cite{Bruni_2013,Leanizbarrutia_2017}, and~other models that assume the presence of \linebreak  condensates~\cite{Cadoni_2018a,Cadoni_2018b,Alexander_2010}; (IV) $f(R)$ theories,~e.g., \cite[]{Sotiriou_and_Faraoni_2010}; (V) mimetic gravity~\cite{Sebastiani_2016}; (VI) a class of models where the cosmic acceleration emerges from the interactions between DM and baryons~\cite{Berezhiani_2017};  (VII) the unified superfluid dark sector~\cite{Ferreira_2019}; (VIII) the fuzzy dark fluid~\cite{Arbey_and_Coupechoux_2021}. 

Among these models, Conformal gravity is one of the first to invoke the unified dark sector, even if it might present some problems in modelling the rotation curves of galaxies and the gravitational lensing effects~\cite{Campigotto_2019}. Two recent and novel models invoking the unified dark sector are the unified superfluid dark sector~\cite{Ferreira_2019} and the fuzzy dark fluid~\cite{Arbey_and_Coupechoux_2021}. In~the unified superfluid dark sector, the~Universe is dominated by a unique DM superfluid made of axion-like particles, with~two energy states having an energy gap smaller than $H_0$ that can interact with each other. These interactions at the microscopic level change the macroscopic behaviour of the fluid, producing an accelerated expansion of the Universe that mimics DE. Besides~the effect of DE, this fluid can also mimic the galaxy phenomenology due to DM without facing some problems encountered by others of these models, such as superluminal sound speeds or the need for a UV~completion.

In the fuzzy dark fluid model, besides~the behaviours of DM and DE, a~single scalar field also mimics the behaviour of inflation, by~assuming a nonminimal coupling to the gravitational field, a~Mexican hat-shape potential, and~a spontaneous symmetry breaking before the inflationary period. This peculiar feature of a unique description of the DM, DE, and~inflation phenomenologies is also shared by mimetic gravity~\cite{Sebastiani_2016}.

\section{Dwarf Galaxies and Globular~Clusters}
\label{sec:Dwarfs_GCs}

The flatness of the rotation curves of disk galaxies and the three scaling relations, BTFR, MDAR, and~RAR, observed on the galaxy scale, are only some of the predictions of MOND. MOND predicted additional pieces of evidence about galactic dynamics before they were observed. Among~them, the~dynamics of LSB galaxies is worth~mentioning. 

As already mentioned in the Introduction (Section~\ref{sec:Intro}), HSB disk galaxies are dominated by stellar mass in their central regions, where their rotation curves are steeply rising toward their asymptotic values. Instead, the~dynamics of LSB galaxies, generally dwarf and dSph galaxies, is observed to be different. Their rotation curves are slowly rising toward their flat region and they appear to be DM-dominated even in their innermost regions. Therefore, the~maximum-disk hypothesis cannot be applied as for HSB galaxies~\cite{Strigari_2008,DiPaolo_2019b}. Specifically, dwarf galaxies are among the known darkest galaxies observed: they have an inner velocity dispersion $\sigma\sim10$~km~s$^{-1}$, an~order of magnitude larger than the velocity dispersion $\sigma\sim1$~km~s$^{-1}$ expected for systems having the same luminosity and scale radius ($\sim$100~pc) at equilibrium~\cite{Mateo_1998}. Their luminosity varies in the range $\sim$[$10^2$, $10^{10}$]~L$_\odot$~\cite{McConnachie_2012,Javanmardi_2016} but their velocity dispersions are similar, which might indicate that they are dominated by a similar DM distribution~\cite{Mateo_1993,deMartino_2020}. 

This different shape of the rotation curves of HSB and LSB galaxies reflects in one of the small-scale issues of the $\Lambda$CDM model, the~cusp/core problem. Indeed, steeply-rising rotation curves can be modelled by a Navarro Frenk White (NFW) DM density \mbox{profile~\cite{Navarro_Frenk_White_1996,Dubinski_and_Carlberg_1991},} cuspy in its innermost part and predicted by collisionless N-body simulations, whereas slowly-rising rotation curves can only be reproduced by a cored DM density profile, which might be accounted for in $\Lambda$CDM only introducing baryonic feedback and tidal effects~\cite{Torrealba_2019}.

The different dynamic properties of these two categories of galaxies were instead predicted by MOND some years before they were observed~\cite{Milgrom_1983b}. Milgrom~\cite{Milgrom_1983b} predicted that dwarf galaxies would have shown strong deviations from standard gravity and in particular that when the velocity dispersion data of dwarf galaxies had been available, these galaxies would have presented a mass discrepancy equal to 10 or larger, depending on their distance from the Milky Way, when modelled in standard~gravity.

As already anticipated in Section~\ref{sec:MOND_RG_BTFR_MDAR_RAR}, the~MOND acceleration scale $a_0$ can be translated on a surface mass density scale $\Sigma_0 = a_0/G$~\cite{Milgrom_1983b,Famaey_and_McGaugh_2012}. Small surface mass densities also indicate small surface brightnesses and Milgrom predicted that LSB galaxies would have shown stronger effects of the modification of the law of gravity with respect to the Newtonian one. The~effect of gravity modification already appears in the innermost radii of these galaxies. Milgrom~\cite{Milgrom_1983b} predicted that a transition radius $r_0$ between the standard and the modified gravity regimes, dependent on the local value of the rotation velocity $V$, would have been set when the equality $V^2/r_0 \thickapprox a_0$ occurs. In~particular, where $V^2/r \gg a_0$, the~local mass-to-light ratio of the galaxy should not indicate the presence of a hidden mass, and~where $V^2/r$ starts to go below $a_0$, the~local $M/L$ should begin to rapidly increase. The~smaller the average surface brightness of the galaxy, the~smaller the $r_0$ in units of the galaxy scale length $h_R$. A~similar result was also found by~\cite{Matsakos_and_Diaferio_2016} for RG (see the top panel of Figure~10 of~\cite{Matsakos_and_Diaferio_2016}). 
Moreover, Milgrom~\cite{Milgrom_1983b} also predicted a correlation between the average surface mass density or surface brightness and the steepness of the rotation curve in reaching its asymptotic limit, which states that galaxies with small surface densities show a slowly-rising rotation curve and vice-versa, as~observed by successive~measurements.

A possible challenge for MOND on galaxy scale is instead provided by the internal dynamics of GCs, which have baryonic masses similar to the ones of LSB galaxies, settling in the outermost regions of the Milky Way. In~those regions, the~background acceleration is much smaller than $a_0$ and the external field effect is negligible. Therefore, we expect that their stellar velocity dispersion profiles present a MONDian behaviour but this is not the case. Newtonian theory without the presence of a DM halo can better fit these kinematic profiles than MOND, whose predicted velocity dispersions can exceed the Newtonian ones also by a factor of $\sim$3~\cite{Baumgardt_2005,Jordi_2009,Baumgardt_2009,Sollima_and_Nipoti_2010,Ibata_2011a,Ibata_2011b,Frank_2012}. Relevant examples for this result are provided by the GCs NGC 2419, Palomar 14, and~Palomar 4~\cite{Jordi_2009,Baumgardt_2009,Ibata_2011a,Ibata_2011b,Frank_2012}. Yet, this tension between MOND predictions and measurements might not indicate an issue of the theory but it might be due to inaccurate data or approximate~modelling. 

Concerning the former case, inaccurate data can derive from low-resolution spectroscopy and from errors on GCs distances larger than 10\% of their values (e.g.,~\cite{Sollima_and_Nipoti_2010}). The~Gaia mission, which provides the parallaxes of the stars from~which to derive their distances, accurate at the $\mu$arcsec level, might represent a turning point in this sense~\cite{Gaia_Collaboration_Helmi_GCs_DwarfGalaxies_2018}. Concerning the latter case, most of the adopted models assume spherical symmetry, absence of rotation, and~orbital isotropy. Whereas the first two assumptions can be justified by observations, orbital anisotropies in GCs are predicted by $N$-body simulations~\cite{Giersz_2006} and including strong radial anisotropies in the modelling can reconcile MOND expectations with the observed velocity dispersions. Further studies with high-precision measurements, which allow us to neatly disentangle the effect of a strong radial anisotropy and the adopted theory of gravity, need to be performed~\cite{Sollima_and_Nipoti_2010}. Moreover, the~question might be even harder to understand, since some theories of formation and evolution of GCs predict the presence of DM in these objects and its observational evidence is debated~\cite{Mashchenko_and_Sills_2005,Moore_1996,Forbes_2008}.

Instead, RG theory provides a more natural solution for the different dynamic properties of LSB, dwarf, and~dSph galaxies and GCs. As~anticipated in Section~\ref{sec:RG_formulation}, RG predicts a different shape for the gravitational field lines in flat and spherical systems. Specifically, whereas in spherical systems the gravitational field lines always maintain a radial direction following the Newtonian trend and become enhanced compared to the Newtonian field in the external regions of the systems, where the density goes below the critical value $\rho_{\rm c}$, in~nonspherical configurations we observe a refraction of the field lines toward the equatorial plane of the object and the deviations from Newtonian gravity already begin to appear in regions where $\rho > \rho_{\rm c}$. In~particular, the~flatter the system the stronger the redirection effect of the field lines and the larger the mass discrepancy if interpreted in a Newtonian framework. This can intuitively explain the diverse dynamic behaviour of LSB, dwarf, and~dSph galaxies and GCs, the~former generally having a flatter shape and the latter a more spherical one~\cite{Matsakos_and_Diaferio_2016}.

This RG prediction is also in agreement with a claimed positive correlation between the elliptical galaxies' ellipticities and their DM content~\cite{Deur_2014,Deur_2020}. As~mentioned in\linebreak   Sections~\ref{sec:Intro} and~\ref{sec:MOND_RG_BTFR_MDAR_RAR}, Cesare~et~al.~\cite{Cesare_2020b,Cesare_2022} demonstrated that RG can model the dynamics of flat (30 disk galaxies from the DMS) and spherical (three E0 galaxies from the SLUGGS survey) systems. The~modelling of the two classes of systems is obtained with statistically consistent $\{\epsilon_0,Q,\rho_{\rm c}\}$ parameters, showing that the gravitational permittivity is independent of the shape of the considered system. To~perform a more robust test of the theory, Cesare~et~al.~\cite{Cesare_2020b} estimated the three RG parameters both from each individual DMS galaxy, simultaneously modelling its rotation curve and vertical velocity dispersion profile, and~from the kinematic profiles of the entire DMS sample considered at the same time. The~two sets of values are consistent within 2$\sigma$. The~average $Q$ and $\rho_{\rm c}$ derived from the simultaneous modelling of the velocity dispersions of the stars and the blue and red GCs in each E0 galaxy are consistent within 1$\sigma$ with the $Q$ and $\rho_{\rm c}$ averaged from the values obtained from each DMS galaxy, whereas the $\epsilon_0$ parameters are still in agreement within 3$\sigma$. The~average $Q$ and $\rho_{\rm c}$ from the E0 galaxies are also consistent, within~3$\sigma$, with~the unique combination of $Q$ and $\rho_{\rm c}$ derived from the entire DMS sample. Instead, the~tension increases to 14.8$\sigma$ for the $\epsilon_0$ parameter. However, this does not necessarily indicate an issue for RG. This might be due to the approximate procedure with which the single combination of RG parameters is estimated from the entire DMS sample, which results in error bars of the $\epsilon_0$ parameter that are much smaller than the error bars of the average $\epsilon_0$ of DMS galaxies. Moreover, it might be due to incorrect modelling of the dynamics of elliptical galaxies, which are treated as isolated systems without net rotation, or~to a wrong assumption for the gravitational permittivity functional form~\cite{Cesare_2022}. A~review of both the works about disk and elliptical galaxies in RG is presented in~\cite{Cesare_2021}.

\section{Discussion and~Conclusions}
\label{sec:Conclusions}

The most investigated cosmological model is $\Lambda$CDM, which assumes the validity of GR and the inclusion of two dark components, DE and DM, besides~baryonic matter, which only represents the $\sim$5\% of the mass-energy budget of the Universe. The~$\Lambda$CDM paradigm reconciles with the majority of the observations, from~the largest to the smallest scales. However, the~results of the detection, through direct, indirect, or~collider experiments, of~the most investigated DM candidate, the~weakly interacting massive particles (WIMPs), are still under debate~\cite{Tanabashi_2018}. Moreover, the~nature of DE is still unknown. Future experiments, such as Euclid, might shed light on its~nature.

Furthermore, the~$\Lambda$CDM model presents some issues, both on large and on small scales. Particularly relevant are the problems observed on the galaxy scale, such as some remarkable coincidences that can hardly be explained by the stochastic merging process of structure formation predicted by $\Lambda$CDM, unless~precise fine-tuning between DM and baryonic processes is invoked. Among~these coincidences, three scaling relations between dark and baryonic matters in galaxies, the~BTFR, the~MDAR, and~the RAR, that neatly quantify the mass discrepancy on galaxy scale, are observed and they see the appearance of the acceleration scale $a_0 = 1.2 \times 10^{-10}$~m~s$^{-2}$. The~results of Mayer~et~al.~\cite{Mayer_2023} from the galaxies in the \textit{Magneticum} simulation can reproduce in $\Lambda$CDM the emergence of this acceleration scale at redshift $z\sim0.1$. They also predicted an evolution of $a_0$ with redshift, where $a_0$ decreases with increasing redshift, which still has to be confirmed. Intriguingly, the~acceleration scale $a_0$ presents another coincidence, being its value consistent with the combination of some cosmological parameters: $a_0\sim H_0\sim\Lambda^{1/2}$. In~particular, the~relation $a_0\sim \Lambda^{1/2}$ links a quantity observed on galaxy scale and the parameter that regulates the Universe accelerated expansion, which suggests a unification of the two dark sectors and connects the physics on small and large scales. This is even less intuitive to interpret in a $\Lambda$CDM~framework.

On the other hand, MOND theory not only reproduces but even predicted, with~either a small or a null intrinsic scatter, these three relations, assuming a modification of the law of gravity for accelerations smaller than $a_0$. The~value of $a_0$ was estimated by Milgrom~\cite{Milgrom_1983b} in several independent ways before it emerged from observations and it turned out to be consistent with the value of $a_0$ observed some years later from the DM-baryons scaling relations. MOND predicted other pieces of evidence on the galaxy scale, such as the fact that LSB galaxies appear ``darker'' and with more slowly-rising rotation curves than HSB galaxies. However, it presents some issues in describing the dynamics of GCs residing in the Milky Way outskirts, which present a Newtonian behaviour even if the background acceleration is below $a_0$. MOND also reproduces the $a_0\sim\Lambda^{1/2}$ relation, as~presented in several studies~\cite{Milgrom_1989,Milgrom_1999,Famaey_and_McGaugh_2012,Milgrom_a0_cosmology_2020}.

A more recent theory of modified gravity, RG, formulated in a nonrelativistic way by Matsakos and Diaferio in 2016~\cite{Matsakos_and_Diaferio_2016}, might shed further light on galaxy dynamics. RG has already presented some encouraging results on galaxy scale, reproducing the dynamics of both flat and spherical systems with a consistent set of RG parameters~\cite{Cesare_2020b,Cesare_2022,Cesare_2021}, the~BTFR, and~the MDAR of real and simulated galaxies. It also models the RAR of DMS galaxies, even if with some issues, which requires further investigations to assess whether these problems are due to the chosen galaxy sample or to RG theory. RG predicts a different shape for the gravitational field lines in spherical and nonspherical systems. Specifically, the~field lines remain radial in spherical systems and become increasingly refracted toward the equatorial plane of increasingly flat systems. The~refraction of the field lines produces a boost of the gravitational field that mimics the presence of a DM halo in Newtonian gravity. This means that an increasingly flat system is increasingly DM-dominated, if~interpreted in the Newtonian context. With~this feature, RG naturally due to the different dynamic properties of LSB galaxies and GCs to their diverse shape, the~former generally being flatter and the latter nearly spherical. 
This system morphology--mass discrepancy relation predicted by RG is also consistent with the possible ellipticity--total M/L correlation of elliptical galaxies estimated by Deur~\cite{Deur_2014,Deur_2020}.

Despite the promising results shown by different theories, we are still far from the answer for a final scenario of the cosmological model. Given the emergence of $a_0$ from several pieces of evidence on galaxy scale, MOND might seem the most intuitive solution. However, it presents several issues on a larger scale. For~example, it can reduce but not eliminate all the mass discrepancy in galaxy clusters~\cite{Sanders_1999,Sanders_2003,Hodson_and_Zhao_2017}. Moreover, the~building of a covariant version of MOND seems to appear challenging. Some attempts at formulating this relativistic extension failed to describe the features of gravitational lenses, provided superluminal speeds, or~were not in agreement with the post-Newtonian tests of General Relativity~\cite{Bekenstein_and_Milgrom_1984,Bekenstein_1988,Sanders_1988}. A~relativistic extension of MOND is TeVeS~\cite{Bekenstein_TeVeS_2004}, which solved some of these problems but was unable to reproduce cosmological pieces of evidence, such as the CMB or the matter power spectra~\cite{Skordis_2009,Bekenstein_2011}. However, further studies about covariant MOND are still ongoing and some recent results might look promising~\cite{Hernandez_2019,Skordis_and_Zlosnik_2021}.

RG might provide an alternative solution. Despite being based on a density scale-dependent modification of the law of gravity, which is not what observations might suggest, a~covariant formulation of RG, CRG, seems to be promising, given the results of~\cite{Sanna_2021} that also show that the acceleration scale $a_0$ might emerge from the WFL of the theory. In~particular, building a relativistic extension of RG was possible since the modification of the law of gravity depends on a scalar quantity, in~this case the density, whereas for MOND it depends on a vector quantity, namely, the~acceleration. CRG describes the DM and DE phenomenologies with a single scalar field, suggesting a unified dark sector, and~retrieves the $a_0\sim\Lambda^{1/2}$ relation, which is a remarkable~result. 

However, further studies have to be performed to validate RG. A~more accurate study of elliptical galaxies, removing the assumptions adopted in~\cite{Cesare_2022} and considering a larger sample with different ellipticities and extended kinematic profiles (e.g., SLUGGS~\cite{Forbes_2017_SLUGGS} and ePN.S~\cite{Pulsoni_2018} surveys), have to be made to better assess if RG can reproduce the dynamics of these systems. The~fact that RG can account for the different dynamics of dwarf galaxies and GCs is only a hypothesis that should be tested on real samples, such as the dwarf galaxies surrounding the Milky Way,~e.g., \cite[]{Salucci_2012} and belonging to the LITTLE THINGS survey~\cite{Oh_2015}, and~the GCs settling in the Milky Way outskirts, in~particular NGC 2419, Palomar 14, and~Palomar 4,~e.g., \cite[]{Baumgardt_2009,Jordi_2009,Ibata_2011b,Frank_2012}. Moreover, RG should be tested on larger scales, to~verify whether it can describe the dynamics of galaxy clusters. Some preliminary encouraging results in this sense were obtained by Matsakos and Diaferio~\cite{Matsakos_and_Diaferio_2016} but these studies have to be extended to larger samples (e.g., CIRS and HeCS~\cite{Rines_and_Diaferio_2006,Rines_2013}) and with less approximate modelling. At~last, the~studies on cosmological scales have to be completed with~CRG.

MOND and RG are only two possible theories of modified gravity but many other theories that might provide a solution for small- and large-scale problems have been built. Another theory that is worth mentioning is scalar-vector-tensor gravity (SVTG), better known as modified gravity (MOG)~\cite{Moffat_2006}. Whereas the modification of the law of gravity in MOND and RG depends on an acceleration and on a density scale, respectively, in~MOG it depends on a length scale. MOG is a theory of gravity built in a covariant way that introduces a scalar, a~tensor, and~a massive vector field, whose contributions are added to the classical EH action. MOG assumes that the gravitational constant $G$, the~coupling constant $\omega$ and the mass $\mu$ of the vector field, and~the cosmological constant $\Lambda$ are dynamical scalar fields which vary with space and time~\cite{Moffat_2006}.

MOG has two progenitor theories, nonsymmetric gravity theory (NGT)~\cite{Moffat_1995} and metric-skew-tensor gravity (MSTG) theory~\cite{Moffat_2005}, which produce the same modified acceleration law as MOG for weak gravitational fields. NGT, MSTG, and~MOG presented several encouraging results on different scales. They can reproduce solar system and terrestrial gravitational tests, the~observations of the binary pulsar PSR 1913+16~\cite{Brownstein_and_Moffat_2006a}, the~rotation curves of both HSB and LSB galaxies and the BTFR~\cite{Moffat_2005,Brownstein_and_Moffat_2006a,Moffat_and_Toth_2009a}, the~dynamics of an elliptical galaxy~\cite{Brownstein_and_Moffat_2006a}, the~velocity dispersion of Milky Way (MW) satellite galaxies~\cite{Moffat_and_Toth_2007a}, the~internal velocity dispersion profiles of GCs~\cite{Moffat_and_Toth_2008}, the~cluster lensing~\cite{Moffat_2005}, the~mass profiles of galaxy clusters derived from $X$-ray emitting gas~\cite{Brownstein_and_Moffat_2006b} and their thermal profiles~\cite{Moffat_and_Toth_2009a},  the~Bullet cluster~\cite{Brownstein_and_Moffat_2007}, and~several cosmological observations, such as the CMB temperature anisotropy, the~galaxy power spectrum, and~the supernova luminosity-distance measurements~\cite{Moffat_2007,Moffat_and_Toth_2007b,Moffat_and_Toth_2009a,Moffat_and_Toth_2009b}. A~more recent work seems to demonstrate that MOG is able to reproduce also the RAR of 153 spiral galaxies from SPARC sample~\cite{Moffat_2016}. 

The dynamics of GCs, of~galaxy clusters, and~of systems at sufficiently large distances from their centres provide an important test to distinguish among MOG, MOND, and~RG. MOG reproduces the internal velocity dispersions of GCs around the MW, independently from their distance from the Galaxy centre, consistently with Newtonian expectations. This is not the case for MOND, which predicts a MONDian behaviour for the dynamics of GCs sufficiently distant from the MW centre, such that the background acceleration goes below $a_0$. However, observations  suggest a Newtonian behaviour also for these GCs, where MOND predictions may exceed the measured velocity dispersions also by a factor of $\sim$3,~e.g., \cite[]{Jordi_2009}. Moreover, as~already specified, whereas MOG can reproduce the masses of galaxy clusters, and~also RG is likely to do this, MOND can only retrieve them with an additional DM~component.

To distinguish among MOG, MOND, and~CRG, it would be ideal to have the kinematic data of different systems at large distances from their centres. Indeed, at~these distances, MOG predicts a Newtonian keplerian trend of the rotation velocity, i.e.,~$V(R)\propto{R^{-1/2}}$, consistent with the results of~\cite{Prada_2003} and with the gravitational lensing results of~\cite{Sheldon_2004}. However, this trend is not predicted by MOND and RG (see Section~\ref{sec:MOND_RG}).

Both MOG and RG, also with its covariant extension CRG, reproduce pieces of evidence on different scales. However, some fundamental features distinguish MOG from CRG. Whereas MOG introduces a scalar, a~tensor, and~a vector fields, CRG only includes a single scalar field. Moreover, whereas the CRG scalar field is responsible for both the DM and the DE phenomenologies, MOG attributes the two dark sectors to two different causes and it introduces a cosmological constant, dependent on space and time, to~mimic DE effect. Concerning the $a_0\sim\Lambda^{1/2}$ relation, whereas CRG retrieves it as a consequence, MOG imposes it~\cite{Brownstein_and_Moffat_2006a} to constrain some MOG parameters ($G_0$, $M_0$, and~$r_0$). Several works,~\linebreak  e.g.,~\cite[]{Brownstein_and_Moffat_2006a,Moffat_and_Toth_2009a}, demonstrate that MOG can model different pieces of evidence with a minimal number or with no free parameters. The~results of Cesare~et~al.~\cite{Cesare_2020b,Cesare_2022} show that a universal combination of RG free parameters might exist. Additional studies have to be performed to verify if the phenomenology at several scales in the Universe can be modelled with this unique combination of RG free parameters. Generally, further tests have to assess which of the three theories of modified gravity, MOND, RG, or~MOG, better describe the pieces of evidence on different~scales.

To conclude, the~correct cosmological model might, at~last, be $\Lambda$CDM and further baryonic processes might still be discovered to properly reconcile the theory with all the observed coincidences on the galaxy~scale.



\vspace{+6pt}
\funding{This research received no external~funding.}

\dataavailability{Not applicable.}

\acknowledgments{I sincerely thank the four referees whose suggestions and corrections improved my review. Part of this work is the result of my Ph. D. activity at the Physics Department of the University of Turin (2017–2021). I sincerely thank all the people that helped me during my Ph. D., in~particular Antonaldo 
 Diaferio, who supervised my Ph. D.~activity.}

\conflictsofinterest{The authors declare no conflict of~interest.} 



\abbreviations{Abbreviations}{
The following abbreviations are used in this manuscript:\\

\noindent 
\begin{tabular}{@{}ll}
AGN & Active Galactic Nuclei\\
BIMOND & bimetric MOND\\
BTFR & baryonic Tully--Fisher relation\\
CEG & Covariant Emergent Gravity\\
CMB & cosmic microwave background\\
CRG & covariant refracted gravity\\
DE & dark energy\\
DM & dark matter\\
DMS & DiskMass Survey\\
dSph & dwarf spheroidal\\
EH & Einstein-Hilbert\\
GCs & globular clusters\\
GR & General Relativity\\
HSB & high surface brightness\\
$\Lambda$CDM & $\Lambda$ cold dark matter\\
LSB & low surface brightness\\
MDAR & mass discrepancy--acceleration relation\\
MOG & MOdified Gravity\\
MOND & MOdified Newtonian Dynamics\\
MW & Milky Way\\
NFW & Navarro Frenk White\\
RAR & radial acceleration relation\\
RG & Refracted Gravity\\
SPARC & Spitzer Photometry and Accurate Rotation Curves\\
SPS & stellar population synthesis\\
SRG & simplified refracted gravity\\
SVTG & Scalar-Vector-Tensor gravity\\
TeVeS & Tensor Vector Scalar gravity\\
WFL & weak field limit\\
WIMPs & weakly interacting massive particles
\end{tabular}
}





\begin{adjustwidth}{-5.0cm}{0cm}
\setenotez{list-name=Note}
\printendnotes[custom] 

\end{adjustwidth}
\newpage
\reftitle{References}

\begin{adjustwidth}{-\extralength}{0cm}
\PublishersNote{}
\end{adjustwidth}
\end{document}